\documentclass[twocolumn,showpacs,preprintnumbers,amsmath,amssymb]{revtex4}

\usepackage[unicode]{hyperref}
\usepackage[caption=false]{subfig}
\usepackage{graphicx}
\graphicspath{ {./images/} }
\usepackage{dcolumn}
\usepackage{bm}
\usepackage[caption=false]{subfig}

\begin{document}

\title{Universality class of explosive percolation in Barab\'{a}si-Albert networks
}%

\author{M. Habib-E-Islam and M. K. Hassan 
}%
\date{\today}%

\affiliation{
University of Dhaka, Department of Physics, Theoretical Physics Group, Dhaka 1000, Bangladesh \\
}

\begin{abstract}
In this work, we study explosive percolation (EP) in Barab\'{a}si-Albert (BA) network,
in which nodes are born with degree $k=m$, for both product rule (PR) and sum rule (SR) of the 
Achlioptas process. For $m=1$ we find that the critical point $t_c=1$ which is the
maximum possible value of the relative link density $t$; Hence we cannot have access to the
other phase like percolation in one dimension. However, for $m>1$ we find that $t_c$ decreases with 
increasing $m$ and the critical exponents $\nu, \alpha, \beta$ and $\gamma$ 
for $m>1$ are found to be independent not only of the value of $m$ but also of PR and SR. It 
implies that they all belong to the same universality class like EP in the Erd\"{o}s-R\'{e}nyi network.
Besides, the critical exponents obey the Rushbrooke inequality in the form $\alpha+2\beta+\gamma=2+\epsilon$ with 
 $0<\epsilon<<1$.

\end{abstract}


\maketitle
Percolation is still studied {\it in extenso} even after more than 60 years
of its first formulation \cite{ref.flory, ref.broadbent, ref.stauffer,ref.saberi,ref.rev_bastas,ref.rev_boccaletti}.
One of the reasons is that its notion is omnipresent in systems as disparate as
spread of forest fire, flow of fluid through porous media, spread of biological 
and computer viruses leading to epidemic, formation of public opinion, resilience of systems etc. 
\cite{ref.Newman_virus, ref.Moore_virus, ref.opinon_1, ref.opinon_2,ref.boccaletti_opinion}.
Another reason is that it can describe  phase transition and critical phenomena albeit it requires no
quantum and many particle interaction effects into consideration \cite{ref.Schwabl}. 
 To study percolation we first need to choose a skeleton. Traditionally, physicists like spatially
embedded lattices as a skeleton. On the other hand, random graph, namely Erd\"{o}s-R\'{e}nyi (ER) network, 
has also been used as a skeleton to study percolation though mostly by mathematicians \cite{ref.erdos}. 
However, use of network as a skeleton has gained huge interest among scientists in general and physicists in 
particular since 1999 when Barab\'{a}si-Albert (BA) proposed their seminal paper on 
scale-free (SF) network \cite{ref.barabasi}. They have revolutionized the notion of the graph theory by showing 
that the probability $F(k)$, that a node in a network is connected to $k$ other nodes, 
decays following a power-law, $F(k)\sim k^{-\lambda}$,  in many real
systems with $\lambda\approx 3$. One of the hallmark features of the SF network is the presence of hubs which is 
hold responsible for making the SF network highly resilient to random removal of links albeit remains vulnerable to 
targeted removal \cite{ref.barabasi_1,ref.pastor}.  The resilience to random attack is in a sense 
opposite to the process of percolation. In percolation one  studies the conditions leading to the 
formation of the giant, a cluster of nodes connected by occupied links that grows linearly with network size $N$, 
while in the resilience one checks the condition of the disappearance of the giant.

In 2009, Achlioptas {\it et al.} proposed a simple variant of the classical random percolation (RP) and 
applied it in the ER network \cite{ref.Achlioptas}. The skeleton of the ER network consists of $N$ 
isolated nodes and 
$N(N-1)/2$ number of frozen links among all the pairs of $N$ nodes. Then at each step
two distinct frozen links are picked at random instead of one which is done in RP. However, ultimately only 
 one of two links, that suppresses the growth of the larger clusters, is activated and occupied 
 while the other link
 is discarded for recycle. A specific rule that discourages the growth of the larger cluster
and encourages the smaller ones to grow faster is the now well-known Achlioptas process (AP). It has been 
found that such a simple change
in the selection rule proved to cause a dramatic effect in the final outcome. 
Indeed, as we keep occupying links following the rules of the AP, we find the emergence 
of a giant cluster with a bang. This is in sharp contrast to its classical counterpart
and hence it is called "explosive percolation" (EP). Indeed, the corresponding order parameter $P$,
the relative size of the largest cluster $s_{{\rm max}}/N$, undergo 
such an abrupt transition that it was at first mistaken as a discontinuity and 
suggested to exhibit the first order or discontinuous transition. This claim immediately
triggered an explosion of scientific activities  \cite{ref.Friedman, ref.ziff_1, ref.fortunato, ref.Costa_2, ref.souza, ref.choi,  ref.ara, ref.da_Costa, ref.Grassberger, ref.Bastas}.
To this end, it is now well settled that the EP transition like its classical counterpart 
is actually continuous but with some non-trivial first order like finite-size effects
\cite{ref.Grassberger, ref.Bastas, ref.Riordan, ref.Choi}.

In this article, we study two variants of EP, namely product rule (PR) and sum rule (SR),
in scale-free BA networks for different $m$ values with which each new node is born. 
We find that the value of $m$ plays a crucial role in determining the value of critical points albeit
$\lambda=3$ independent of $m$. First, we show that for $m=1$, the critical point $t_c=1$ 
which is the maximum value of $t$ and hence like one dimensional percolation model
we do not have access to the $t>t_c$ phase. Second, we find that the critical points decrease 
with increasing $m$.  Despite more than $60$ years long history of extensive studies the definition of entropy for percolation remained elusive  until our recent works \cite{ref.hassan_didar,ref.hassan_sabbir}. 
Note that not knowing the entropy makes percolation as a model for phase transition incomplete. 
It is even more the case for a model like EP as it is originally 
claimed to describe the first order transition. The ultimate litmus test of first order transition is 
the existence of latent heat which is connected to the size of the gap in entropy at $t_c$. 
Third, we show that not only entropy is continuous across $t_c$ but it is also consistent with the
behaviour of the order parameter which
suggests that percolation  is actually accompanied by order-disorder transition.
Besides this, knowing the entropy paves the way for defining specific heat. Recently, we have also used jump 
in the order parameter per unit step as the susceptibility. Fourth, using these definitions and applying the
idea of data-collapse we obtain the critical exponents $\nu, \alpha, \beta, \gamma$ and  find 
that both PR and SR share the same critical exponents independent of $m$ provided
$m>1$.  Finally, we show that the critical exponents of this unique universality 
class obey the Rushbrooke inequality in the form of $\alpha+2\beta+\gamma=2+\epsilon$ with $0<\epsilon<<1$.

We begin by discussing the construction process of the BA network since this is the skeleton
we have used in this work. We first choose a seed consisting of $m_0$ 
arbitrarily connected nodes where $m_0$ has to be extremely small compared to the final size $N$
of the network that we grow. Most of the observable quantities of the BA network, however, are independent
of the choice of the size of the seed and of the detailed nature of how seed nodes are connected. 
Once a seed is chosen we keep adding one node with $m$ links at each time step where $m\leq m_0$. The new node
then picks $m$ distinct existing node following preferential attachment probability $\Pi(i)\propto k_i$
that embodies the intuitive idea of rich get richer mechanism. Incorporating (i) kinetics and
(ii) preferential attachment (PA) rule, BA showed
that the resulting network exhibits power-law degree distribution $P(k)\sim k^{-\lambda}$ where $\lambda=3$.
The BA networks of size $N$ consist
approximately of $mN$ links or bonds since each node is born with degree $k=m$. Each
link contributes to degree $2$. Thus the total degree is equal to $2mN$ and hence the average degree of the BA network of size $N$ is $2m$.
Network grown in this way can be used as a skeleton provided its properties are time  
{\it vis-a-vis} the size independent. We have shown in Ref. \cite{ref.hassan_dynamic_scaling} that as the network grows, its
snapshots at different times are statistically similar in the same sense as two triangles are similar if the
numerical values of their dimensional quantities differ but those of the corresponding dimensionless quantities
coincide. 
Since the snapshots of the same system at different times are similar we can say that BA network
exhibits temporal self-similarity.

Once the network of desired size is grown we use that as a skeleton like 
we earlier used scale-free weighted planar stochastic lattice and ER network \cite{ref.hassan_sabbir,ref.hassan_dynamic_scaling, ref.hassan_EP_wpsl}. 
We assume that all the links of the BA network are temporarily frozen.  We label the frozen links as
 $e(i,j)$ which represents a link that establishes a connection between nodes $i$ and  $j$ 
such that $i\neq j$ since self connection is forbidden. 
Earlier, two groups independently studied EP to form scale-free networks without growth
\cite{ref.fortunato, ref.choi}. In their works, they started with a fixed number of nodes and a given sequence
of degree  or weight. Then at each step a pair of candidate links are picked
preferentially with respect to degree  or weight. In contrast, we use
the scale-free BA network as a skeleton from where we pick a pair of candidate links randomly 
at each step and follow the rest of the process exactly like Achlioptas {\it et. al}  \cite{ref.Achlioptas}.
 The process of occupying in such fashion goes on until 
$n=N$ links are occupied or $t=n/N=1$ is achieved.
Links can be of two types. When occupation of a link connects two isolated clusters then it results 
in a larger cluster and it is called inter cluster link. On the other hand, when a link connects two sites 
of the same cluster, which does not result in the increase in the cluster size, then the corresponding link
is called intra-cluster link.
Initially, every node is a cluster of its own size since all the links are assumed frozen.
The process starts by picking two distinct links, say $e(i,j)$ and $e(k,l)$, randomly with 
uniform probability at each step. 
To apply the PR, we then calculate the products, $\Pi_{ij}=s_i\times s_j$ and 
$\Pi_{kl}=s_k \times s_l$,
of the size of the clusters that the two nodes on either end of the links $e(i,j)$ and $e(k,l)$
respectively contain. The link with the smaller value of the products $\Pi_{ij}$ and $\Pi_{kl}$
 is occupied; in case of $\Pi_{ij}=\Pi_{kl}$ one of the corresponding links is selected randomly.
On the other hand, to apply the SR, we take the sum $\Sigma_{ij}=s_i+s_j$ 
and $\Sigma_{kl}=s_k+s_l$ instead of the product and do the same as for PR.

Entropy $H$ and order parameter $P$ definitely are the two most important quantities of interest in
the theory of phase transition as they are used to define the order of transition. For instance, if they
suffer a sudden jump or discontinuity at the critical point then the transition 
is first order and else it is called continuous or second order phase transition. 
Besides, they are also used as a litmus test
of whether the transition is accompanied by symmetry breaking or not. In the case of symmetry breaking
the system undergoes a transition from the disordered state, characterized by maximally high
entropy, to the ordered state, characterized by maximally high order parameter.
Note that the highly ordered state is always less symmetric than the highly disordered state. 
For instance, in ferromagnetic to paramagnetic transition we find that  $P$ and $H$
undergo such an abrupt or sudden change but without gap or discontinuity at $t_c$. 
In the case of percolation, it has been well known that the order parameter changes in the same
fashion as in ferromagnetic transition. However, until we how entropy changes we cannot know whether
the percolation transition too is accompanied by symmetry breaking. 
Only recently we proposed a suitable normalized probability and
obtained desired behaviour of entropy for the first time by using it in the definition of Shannon entropy  
\begin{equation}
\label{eq:shannon_entropy}
H(t)=-K\sum_i^m \mu_i\log \mu_i,
\end{equation} 
where we choose $K=1$ since it merely amounts to a choice of a unit of measure of entropy 
\cite{ref.shannon}. 
Surely one can use any normalized probability in Eq. (\ref{eq:shannon_entropy})
and claim to have measured entropy. Especially, being percolation a probabilistic model, there is no short of
normalized probability. However, note that Eq. (\ref{eq:shannon_entropy}) is not over size rather
over sequence of labeled probabilities. In fact, all the normalized probabilities that exist in percolation
can only be used if the sum is over cluster size $s$ not over individual labeled cluster, yet
authors used them to measure entropy \cite{ref.Vieira, ref.Tsang}.

\begin{figure}

\centering

\subfloat[]
{
\includegraphics[height=2.4 cm, width=4.0 cm, clip=true]
{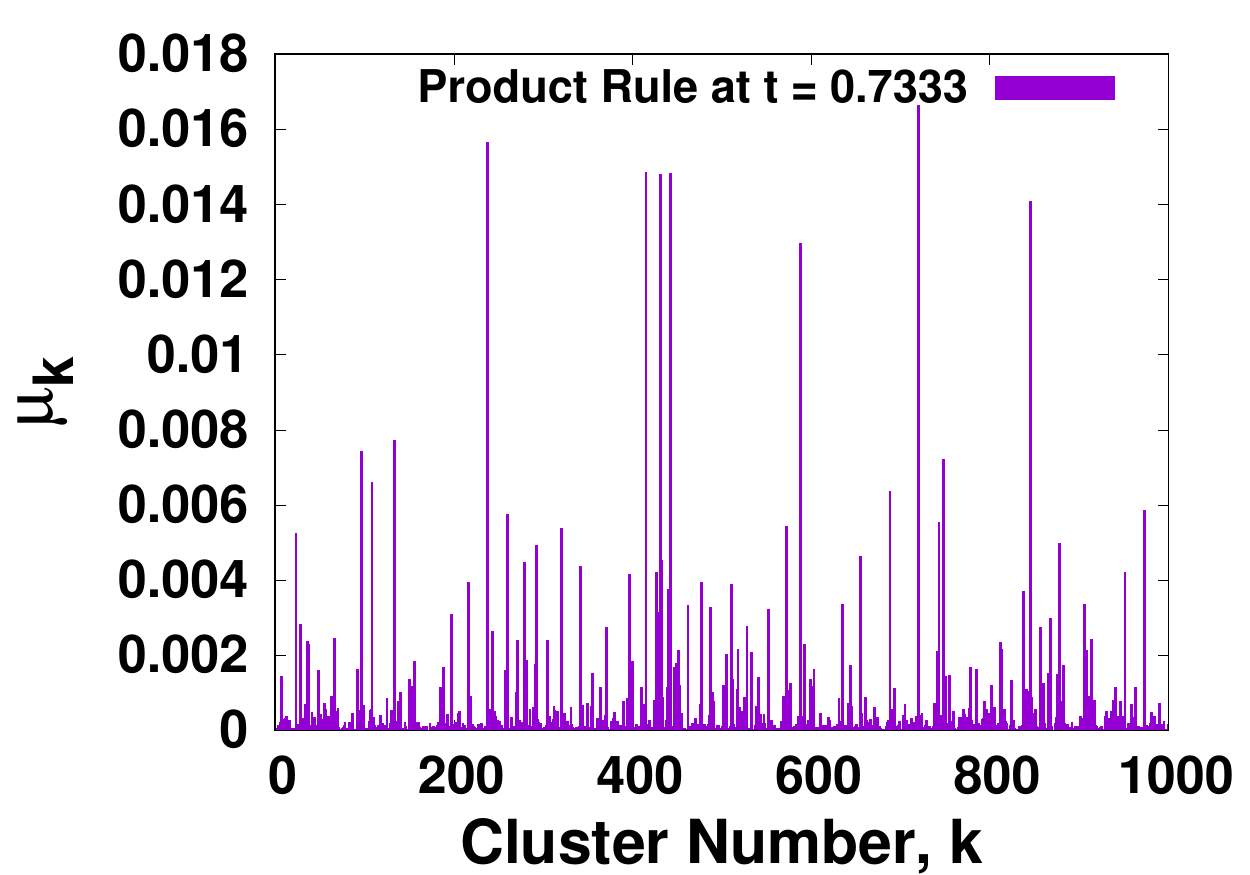}
\label{fig:1a}
}
\subfloat[]
{
\includegraphics[height=2.4 cm, width=4.0 cm, clip=true]
{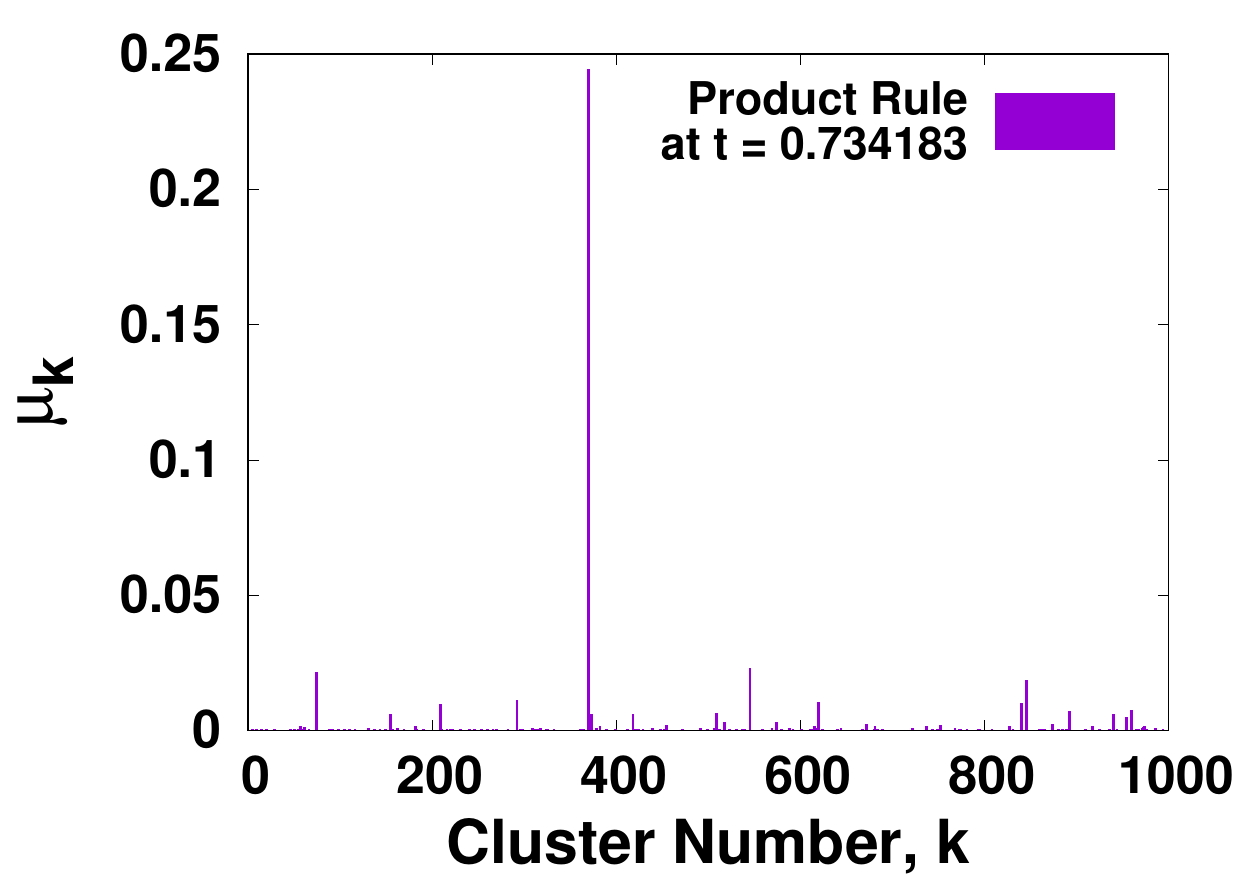}
\label{fig:1b}
}

\caption{Plots of cluster picking probability $\mu$ as a function of cluster label $i$ to 
show abruptness of its change as $t$ changes by a negligibly small amount.
} 
\label{fig:1ab}
\end{figure}

In order to introduce entropy for EP, we consider that
there are $m$ distinct, disjoint, and indivisible labelled clusters $i=1,2,...,k$ 
whose sizes are say $s_1,s_2,....,s_k$ respectively  after addition of $n=tN$ links. 
The Shannon entropy was introduced in the context of information theory to
measure the information contained in each message in a given flow of information. 
By regarding each cluster as equivalent to a message we have recently defined
a set of cluster picking probabilities $\{\mu_1, \mu_2,...,\mu_k\}$ where $\mu_i=s_i/N$ 
quantifies how likely that a site being picked at random belong to the cluster $i$
 \cite{ref.hassan_didar, ref.Hassan_Rahman_1}. Note that initially each node is a cluster of its 
own size $s=1$ and hence $\mu_i=1/N$ for all the nodes $i=1,2,...,N$. 
This is exactly like the state of the isolated ideal gas where all allowed microstates are 
equally likely and hence it is expected that entropy is maximum at $t=0$. Before we
see what happens at the other extreme, at $t=1$, we first plot histogram of $\mu_i$ versus 
cluster label $i$ in Fig. (\ref{fig:1ab}) for PR with $m=5$ only to see what happens to the probabilities 
$\{\mu_1, \mu_2,...,\mu_k\}$ in the vicinity of $t_c$. The same is true for SR and for other values
of $m$ as well. It is clear from the
figure that already near $t_c$ there is a clear sign of explosive growth to one giant cluster at $t_c$. 
Thus, it is expected that at $t=1$  almost all the nodes belong to one giant cluster making 
$\mu\approx 1$ and hence it is expected that entropy at $t=1$ is minimally low.

\begin{figure}

\centering

\subfloat[]
{
\includegraphics[height=2.4 cm, width=4.0 cm, clip=true]
{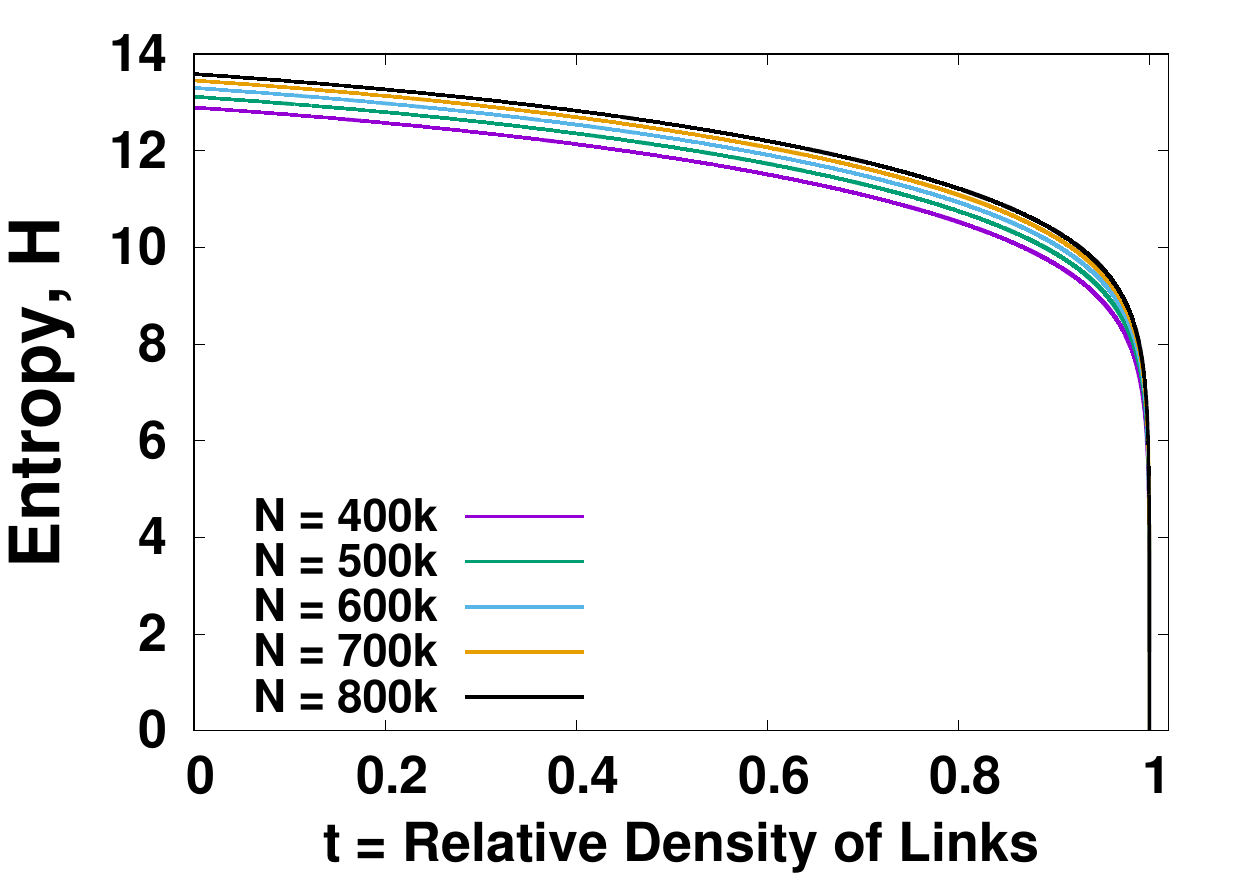}
\label{fig:2a}
}
\subfloat[]
{
\includegraphics[height=2.4 cm, width=4.0 cm, clip=true]
{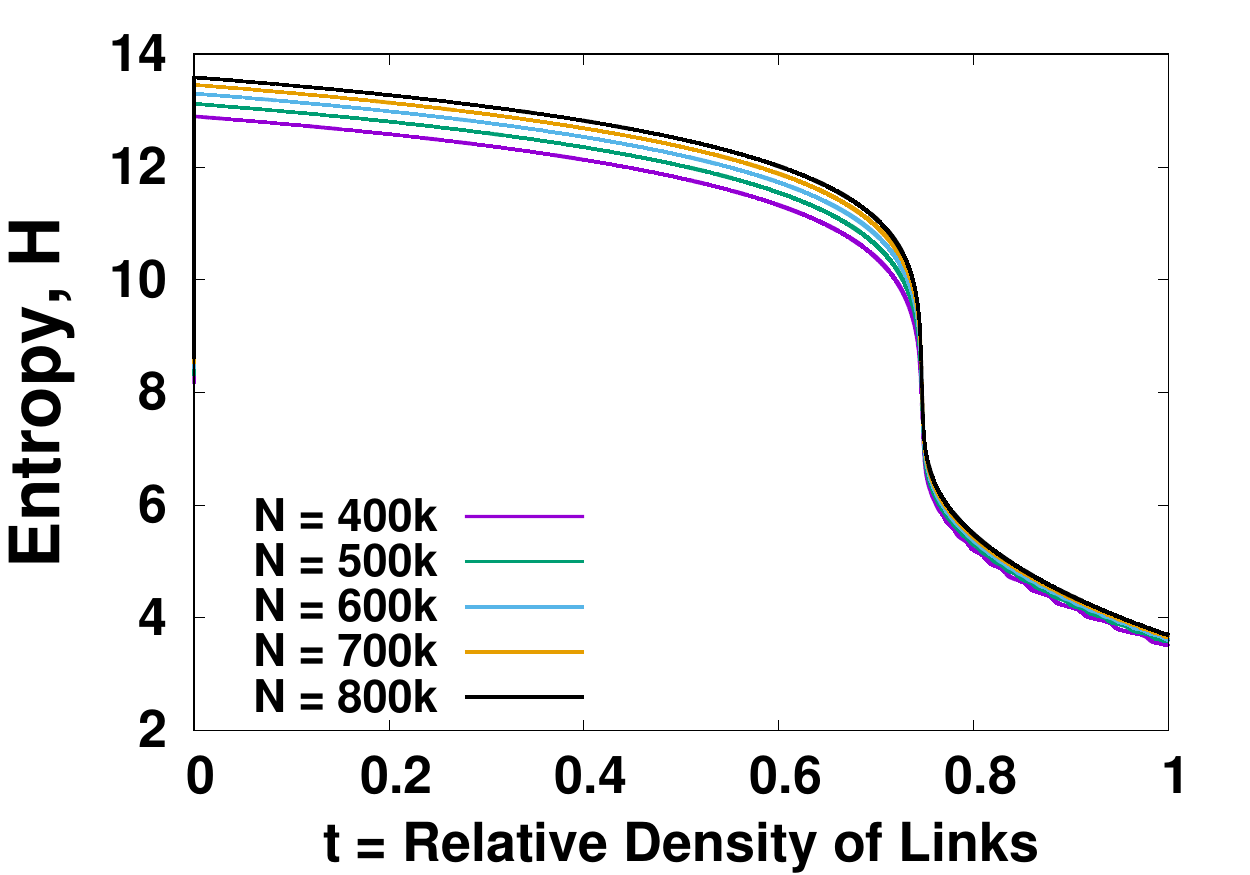}
\label{fig:2b}
}

\subfloat[]
{
\includegraphics[height=2.4 cm, width=4.0 cm, clip=true]
{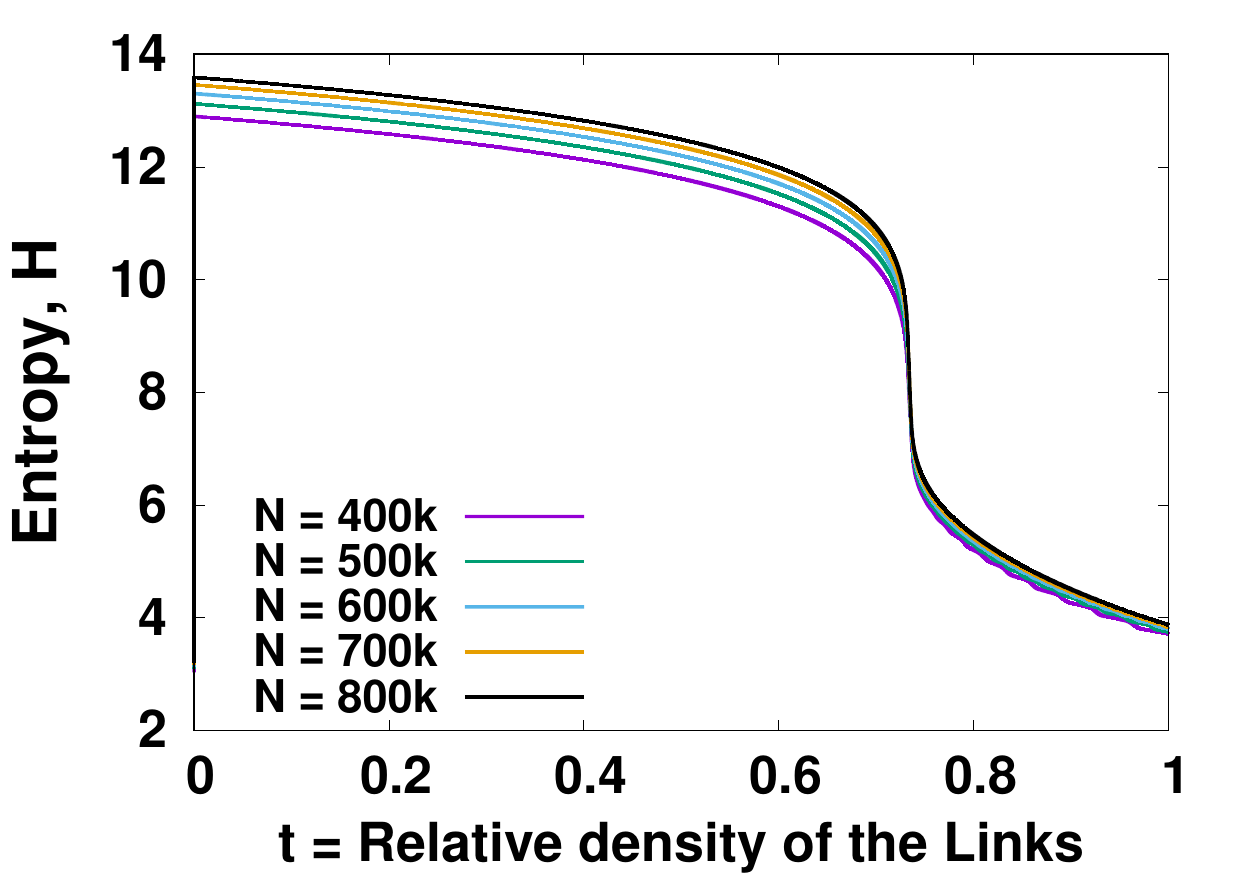}
\label{fig:2c}
}
\subfloat[]
{
\includegraphics[height=2.4 cm, width=4.0 cm, clip=true]
{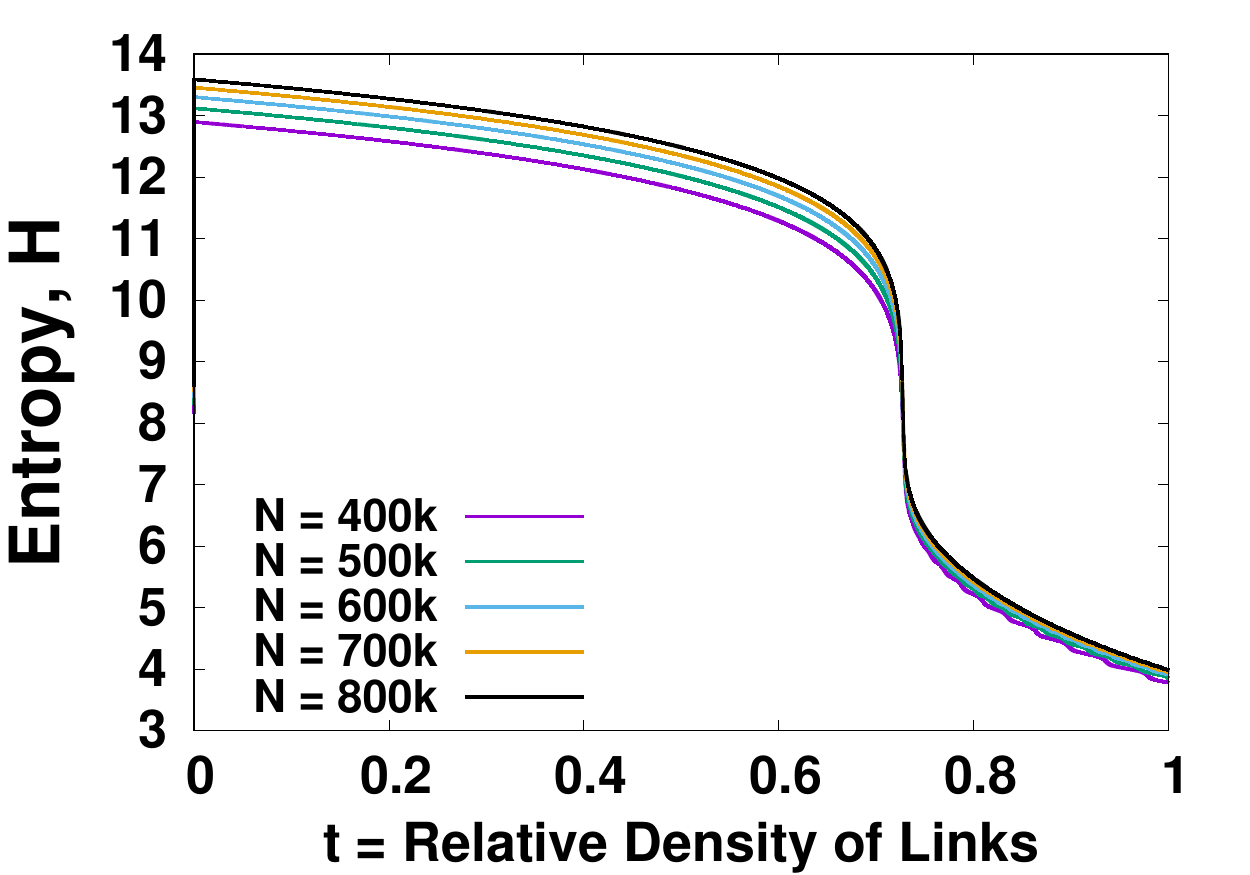}
\label{fig:2d}
}

\caption{ Plots entropy $H$ vs $t$ for $m=1,3,5$ and $m=10$. We 
find the critical points for $m=1,3,5$ and $m=10$ are $1, 0.7467,0.73422$ and $0.7272$ 
respectively. These plots clearly reveals that $t_c$ values decreases with increasing $m$ but they all 
except $m=1$ share the same shape.
} 

\label{fig:2abcd}
\end{figure}

Using the cluster picking probability $\mu_i$ 
in Eq. (\ref{eq:shannon_entropy}) we have measured entropy which is shown in Fig.
(\ref{fig:2abcd}) for a few different $m$ values.  We see that for $m=1$ the critical point $t_c=1$. This
 is like one dimensional percolation problem where we do not have access to the other phase
$t>t_c$ since the number of links $n$ for $m=1$ cannot exceed $N$ and hence $t\leq 1$.
However, as we increase the value of
$m$ from $m=1$ we find that $t_c<1$ and the $t_c$ values decreases with increasing $m$. Therefore, we
will only concentrate on $m>1$ case in the rest of the article. We can see from Fig. (\ref{eq:shannon_entropy}) 
that the entropy versus $t$ curves interpolate nicely between $t=0$ and $t=1$. We all know entropy
measures the degree of disorder. What is disorder in percolation anyway? Say initially every
cluster is colored with a distinct color. To do that we need $N$ number different colors.
The probability that a node picked at random belongs to, say green
or red or any other color, is $1/N$. Seeing a system with number of different
colors as many as the number of clusters will definitely give an impression of a disordered
system. Now as we keep adding links, clusters will merge to form bigger clusters. Consider
that when two clusters merge the resulting cluster take the color of the larger cluster and in the
case of merging  two equal sized clusters the resulting picks one of the colors at random. As we
continue adding links we will see a transition across $t_c$ to a giant cluster of one single color 
that dominates the entire system. 
At this point, the system will look ordered in the sense that any site we pick at random is almost certain
that it will belong to the giant cluster.  


\begin{figure}

\centering

\subfloat[]
{
\includegraphics[height=2.4 cm, width=4.00 cm, clip=true]
{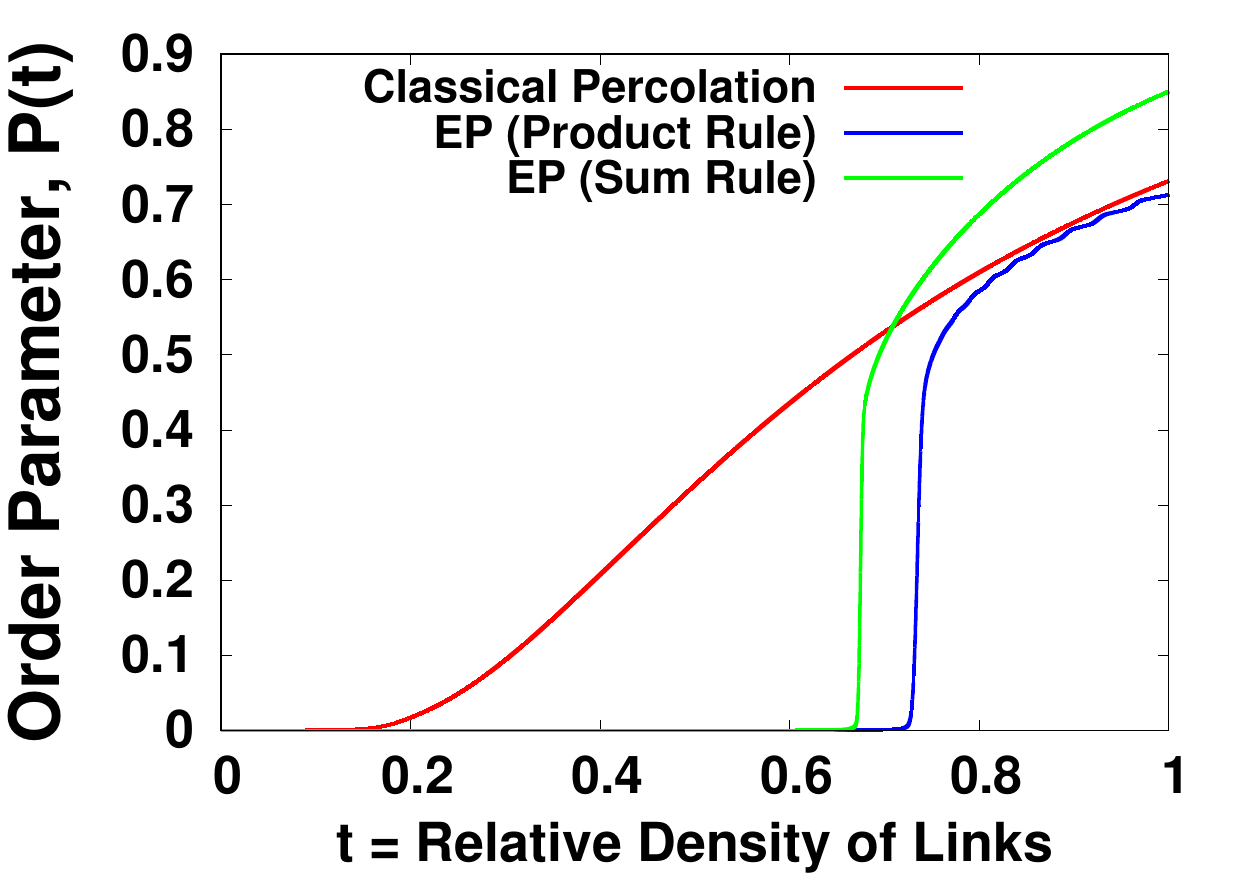}
\label{fig:3a}
}
\subfloat[]
{
\includegraphics[height=2.4 cm, width=4.00 cm, clip=true]
{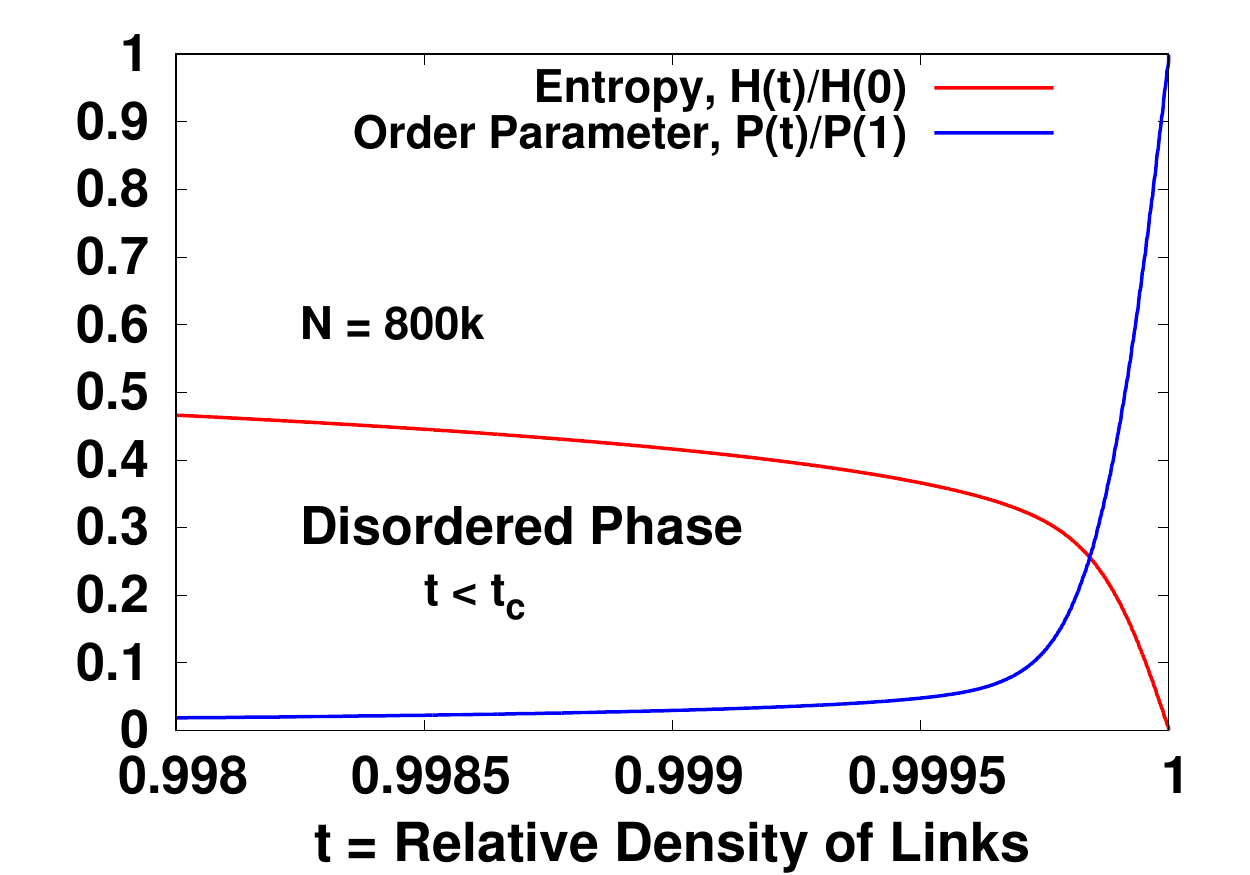}
\label{fig:3b}
}

\subfloat[]
{
\includegraphics[height=2.4 cm, width=4.00 cm, clip=true]
{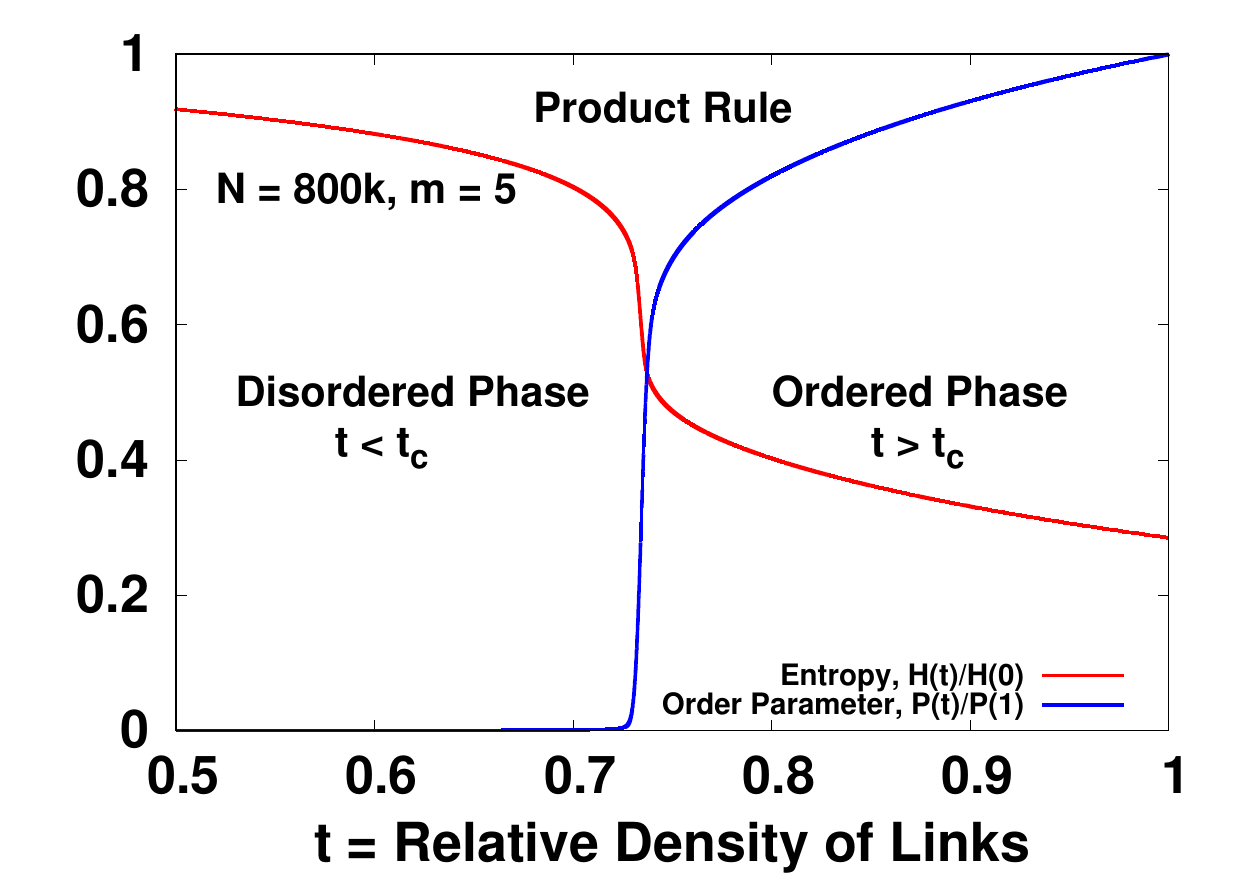}
\label{fig:3c}
}
\subfloat[]
{
\includegraphics[height=2.4 cm, width=4.00 cm, clip=true]
{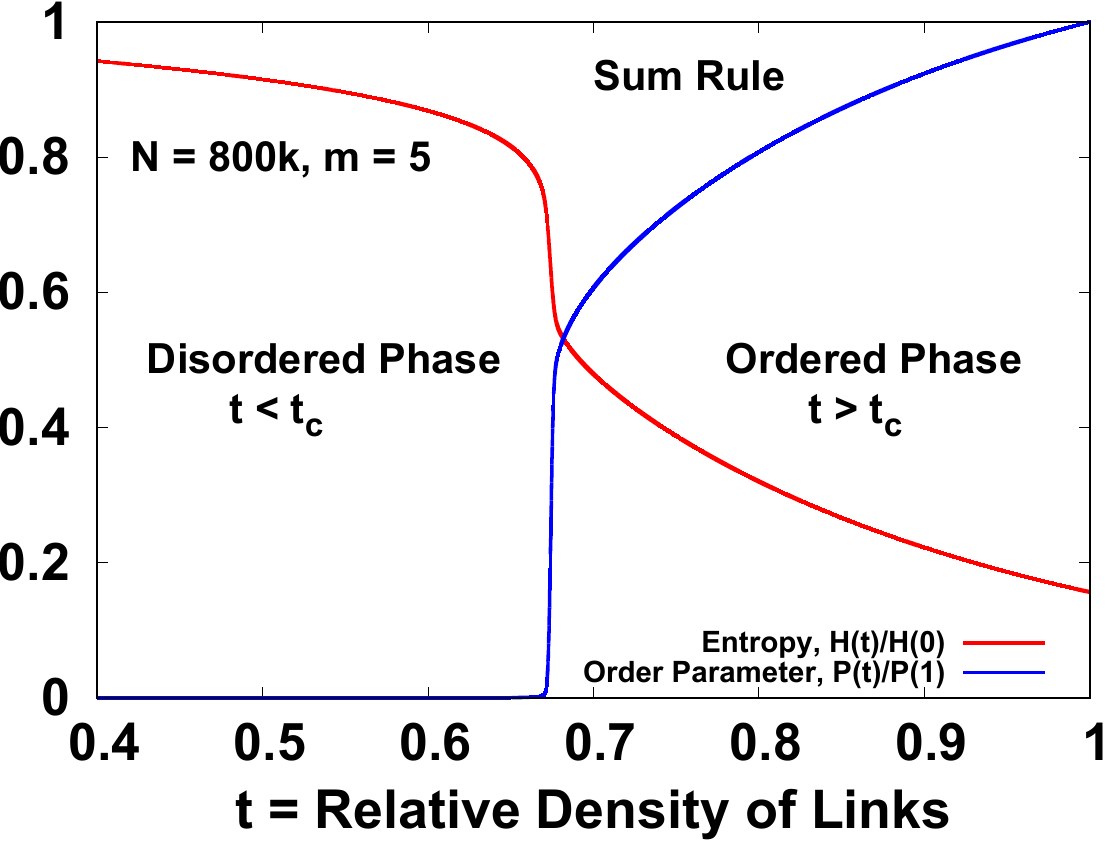}
\label{fig:3d}
}
\caption{(a) Comparison of order parameter for random percolation and explosive percolation (PR and SR) in the 
BA network. In (b)-(d) we show plots of relative entropy $H(t)/H(0)$ and relative
order parameter $P(t)/P(1)$ for different $m$ and for different rules. In  
(b) we plot for $m=1$ that clearly shows $t_c=1$ and we plot the same for $m>1$ in (c) for PR and
(d) for SR. 
} 
\label{fig:3abcd}
\end{figure}

Indeed, the relative size of the largest cluster $P=s_{{\rm max}}/N$  is
defined as the order parameter. Note that at $t\leq t_c$ the largest cluster 
is miniscule in size which decreases in comparison to $N$ as $N$ increases and hence
 $P\rightarrow 0$ as $N\rightarrow \infty$ till $t=t_c$. However,
at above $t_c$ the largest cluster is almost as large as the size of the system
$N$ itself. Thus, in the $t> t_c$ regime 
$P$ is non-zero and for a given value of $t$ its value increases to a constant as 
$N\rightarrow \infty$. Moreover, $P$ increases sharply near $t_c$ and then moves slowly
towards unity as $t$ increases further. This is exactly how magnetization behaves with temperature
during the ferromagnetic to paramagnetic phase and hence regarding $P$ as the order parameter is well justified. 
In Fig. (\ref{fig:3a}) we show $P$ for random and explosive percolation in BA network. We see
that $t_c$ for RP in the BA network is significantly smaller compared to that for the RP in the ER network. Similarly,
$t_c$ value of EP in the BA network is also significantly smaller compared to that for the EP in the ER network
\cite{ref.Achlioptas}. Besides, the sudden and abrupt growth of $P$
which is typical to EP is absent in the RP. Yet,  another point to note is that the extent of rise
of $P$ in the case of EP in network, regardless of whether it is in the BA or ER network, is much too sharp
than the same in the lattice $t_c$ \cite{ref.Hassan_Rahman_explosive}. Our extensive
investigation of RP in the BA and ER network suggests that RP 
is more appropriate in the lattice than in the network. 
Note that we tried to find the various critical exponents for RP in the BA network but it went in vain. 

Typically, the phase where the order parameter is always
zero is regarded as the disordered phase provided entropy is significantly high there. On the other hand,
the phase where entropy is minimally low but the order parameter is significantly high is called the ordered phase. In Fig. (\ref{fig:3abcd}) we present both the quantities in the same plot to see whether
they compliment each other in this fashion or not. 
However, note that the numerical value of the maximum entropy, which is equal to $\log (N)$,
is much higher than that of the $P$, which can at best be equal to unity. Therefore,
for better comparison we plot relative entropy $H(t)/H(0)$ and relative order parameter $P(t)/P(1)$ 
in an attempt to re-scale their values so that in either cases their respective maximum values become
unity. The plots of re-scaled entropy and order parameter
are shown in Figs. (\ref{fig:3b})-(\ref{fig:3d}) for various $m$ values and for the two rules under
investigation.
These figures clearly show that when $P=0$, the entropy $H$ is maximally high and
vice versa. It means that percolation transition, like paramagnetic to ferromagnetic transition, 
is accompanied by order-disorder transition.

\begin{figure}

\centering

\subfloat[]
{
\includegraphics[height=2.4 cm, width=4.00 cm, clip=true]
{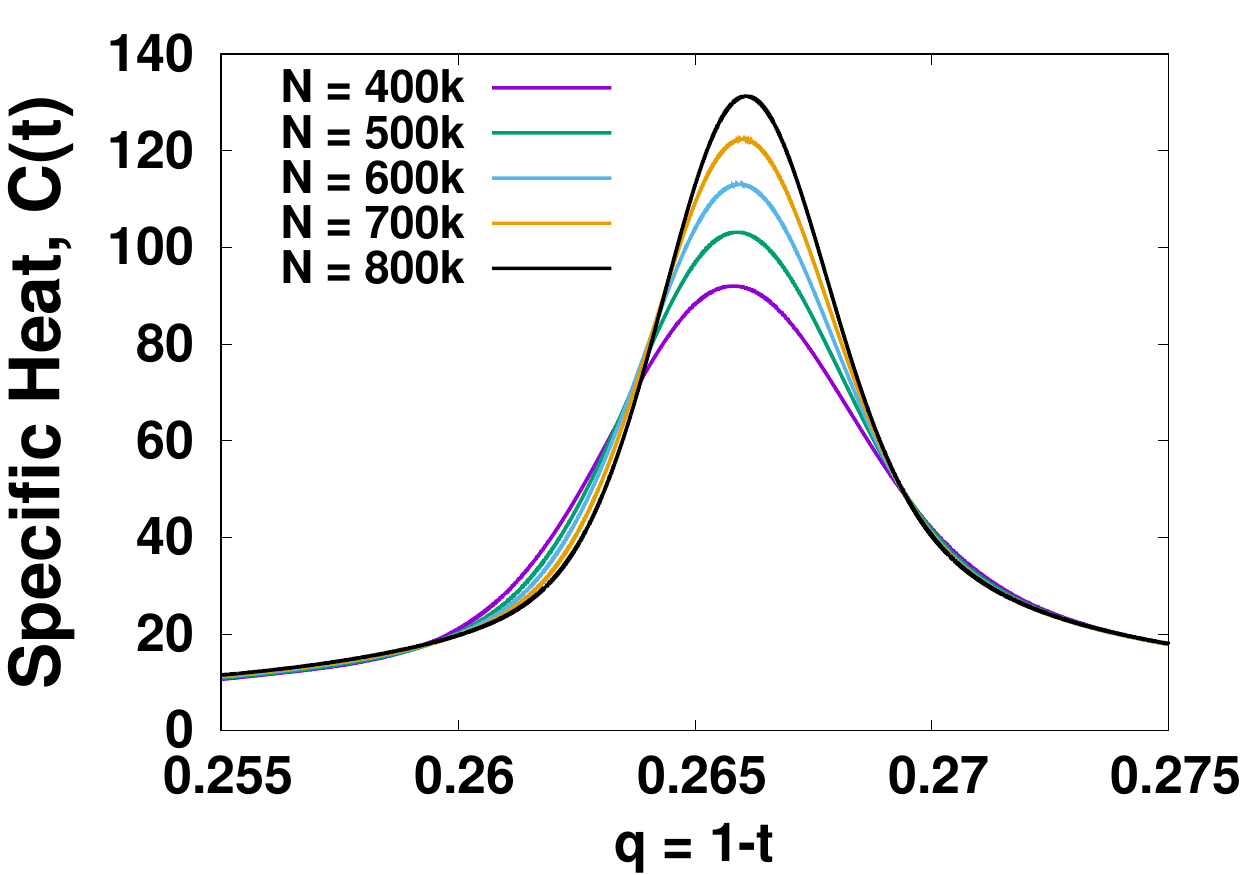}
\label{fig:4a}
}
\subfloat[]
{
\includegraphics[height=2.4 cm, width=4.00 cm, clip=true]
{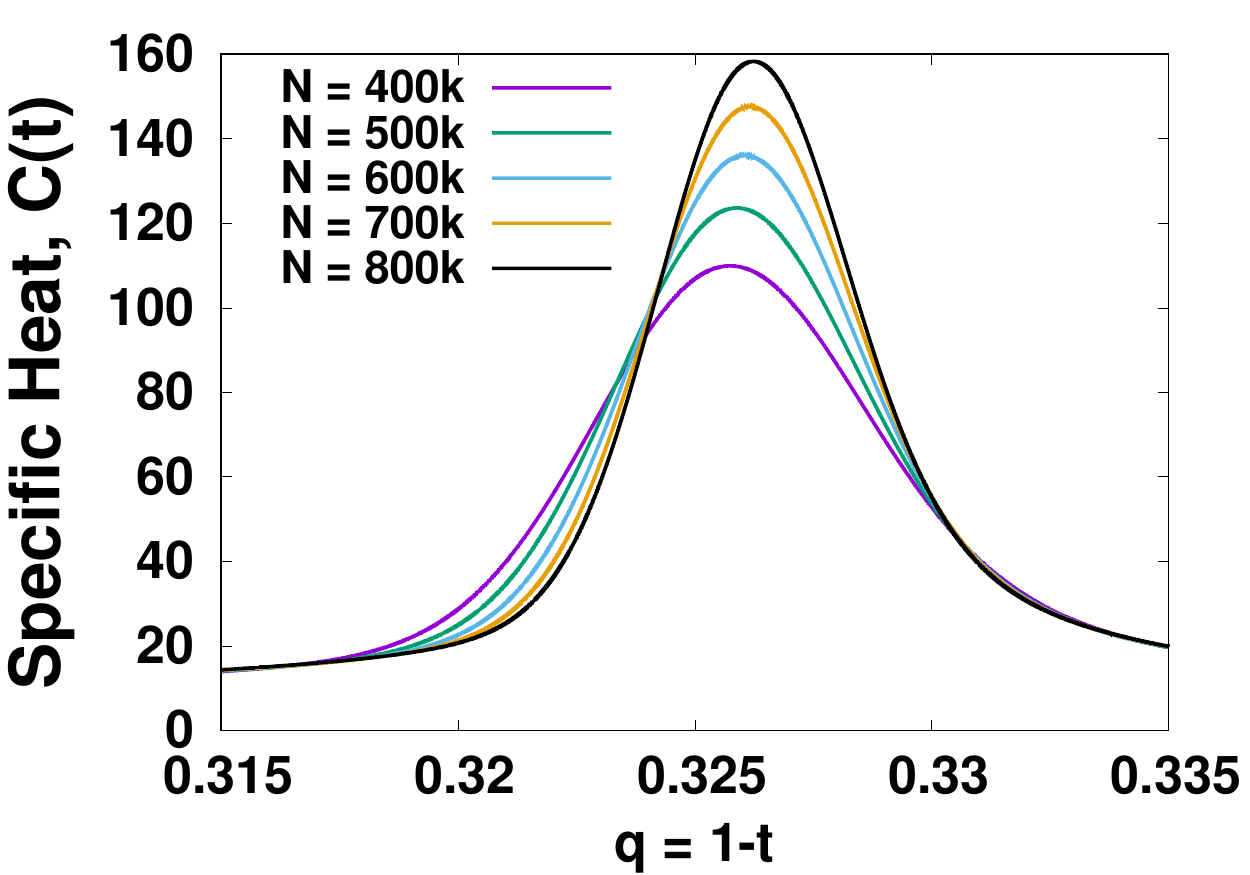}
\label{fig:4b}
}

\subfloat[]
{
\includegraphics[height=2.4 cm, width=4.00 cm, clip=true]
{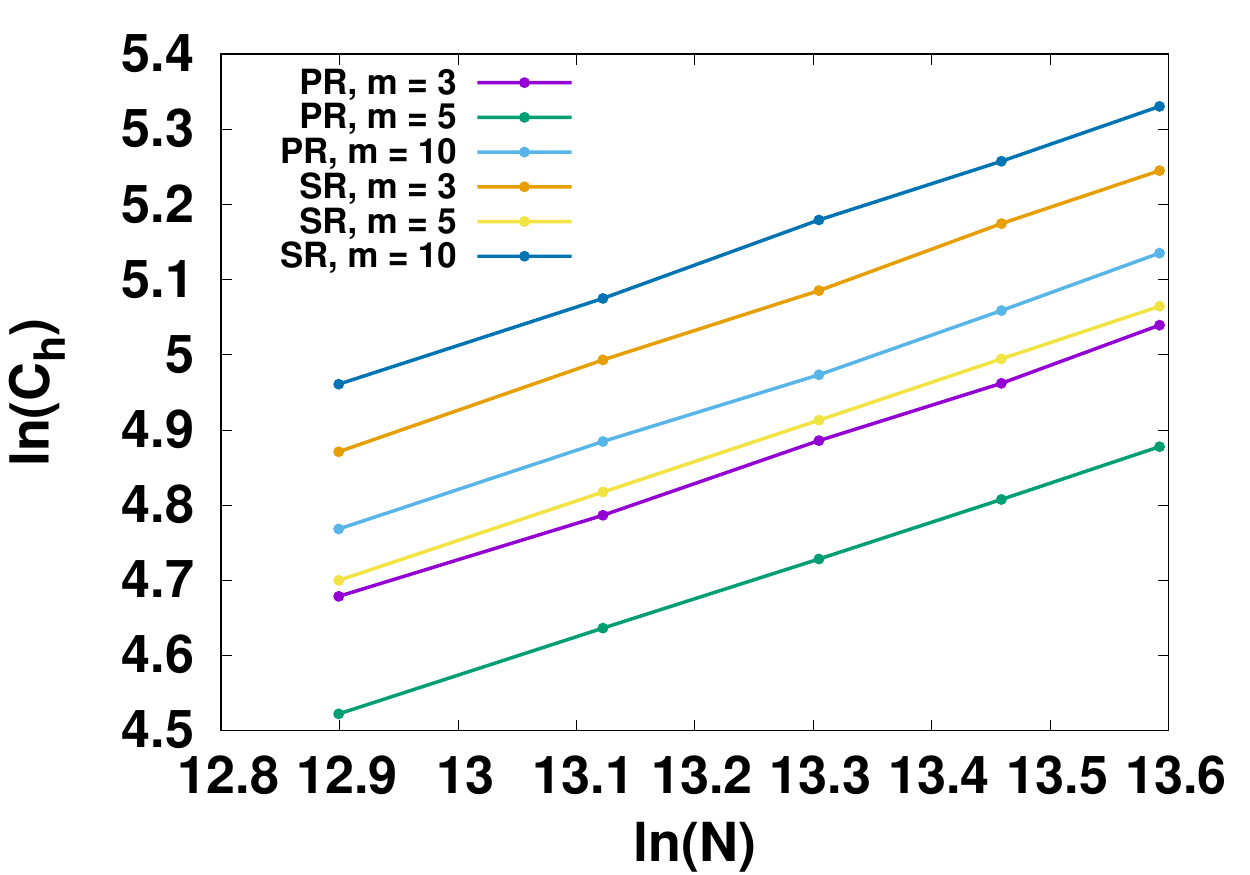}
\label{fig:4c}
}
\subfloat[]
{
\includegraphics[height=2.4 cm, width=4.00 cm, clip=true]
{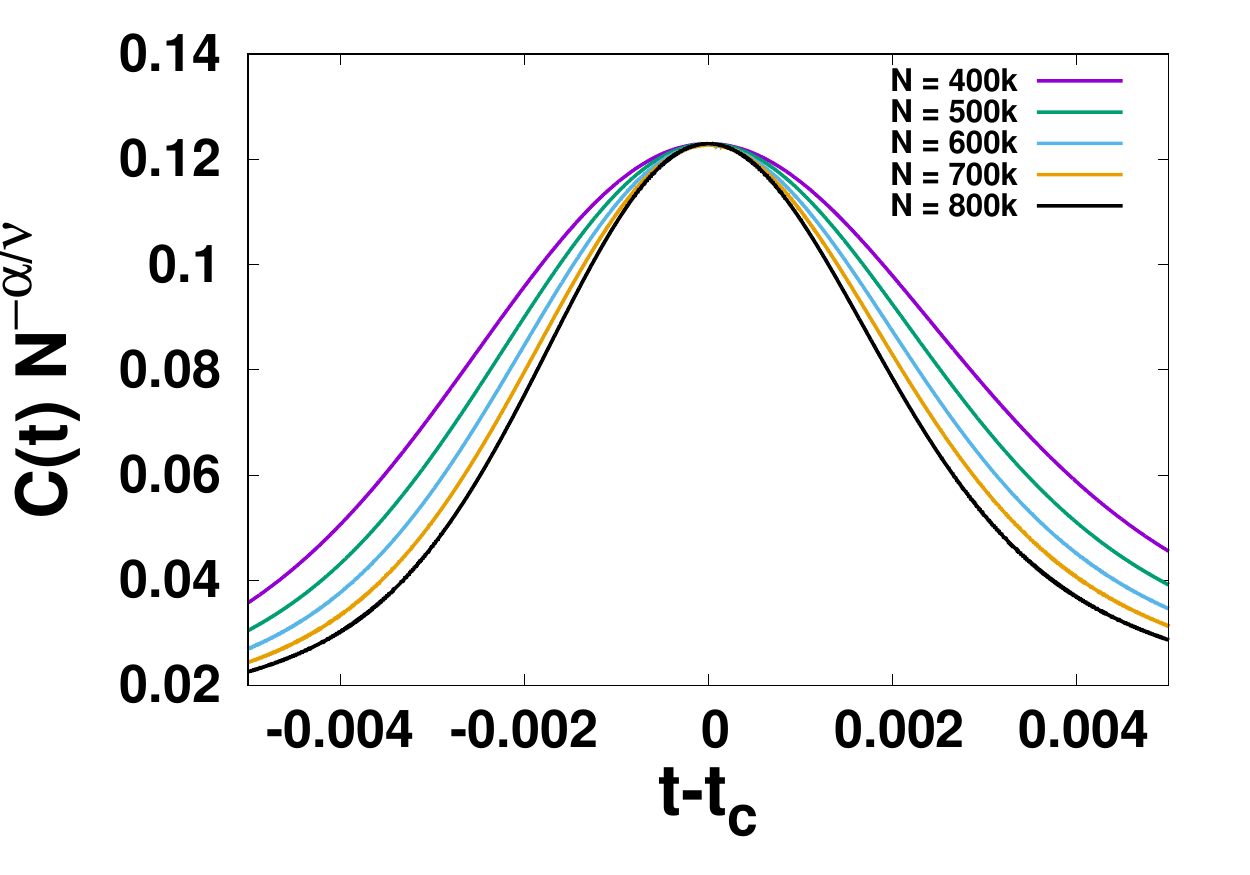}
\label{fig:4d}
}

\subfloat[]
{
\includegraphics[height=2.4 cm, width=4.00 cm, clip=true]
{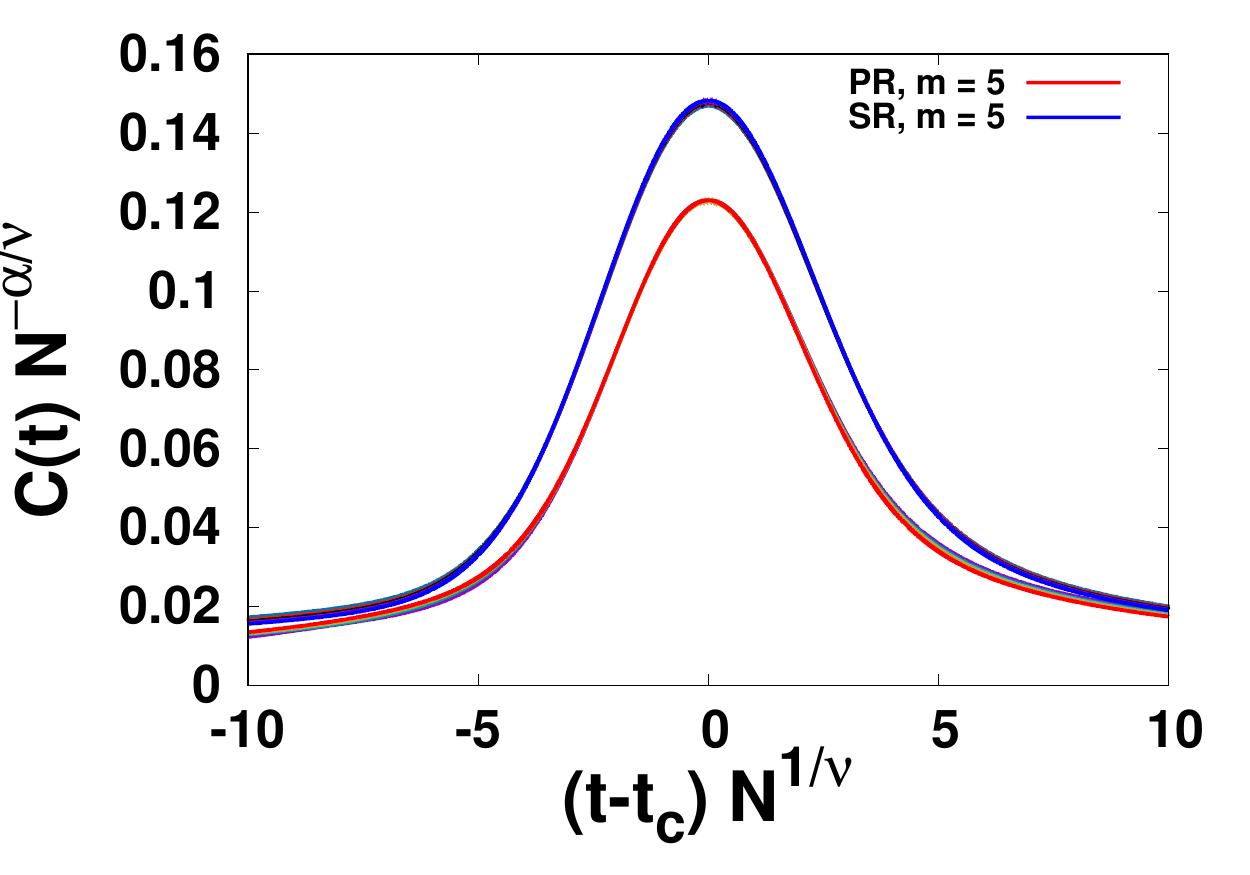}
\label{fig:4e}
}
\subfloat[]
{
\includegraphics[height=2.4 cm, width=4.00 cm, clip=true]
{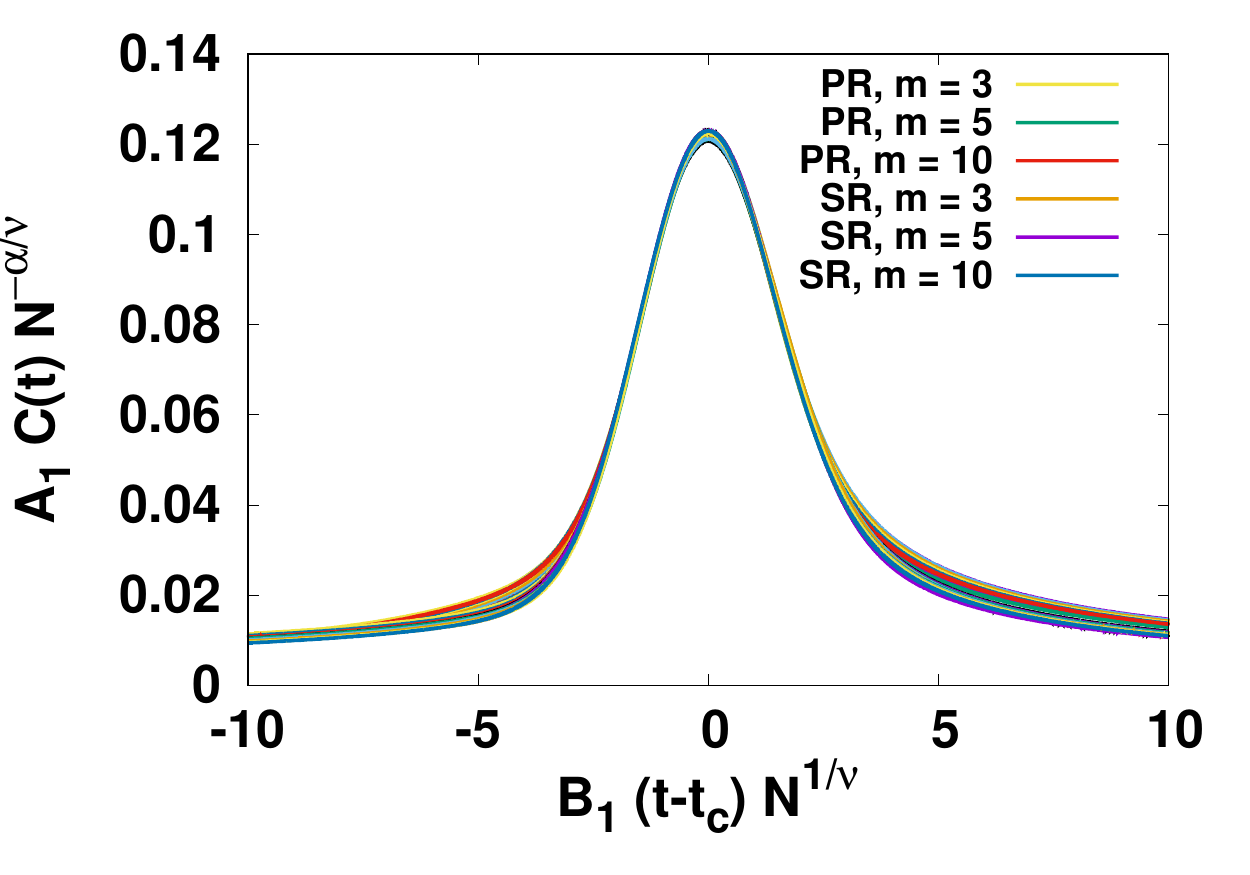}
\label{fig:4f}
}
\caption{
Specific heat $C$ versus $t$ for (a) PR and (b) SR in the BA network with $m=5$. (c) The slopes of the plot
$\log(C_{{\rm h}})$ vs $\log(N)$, where $C_{{\rm h}}$ is the peak value at $t_c$,
 for different $m$ and different rules gives values of $\alpha/\nu$. In (d) 
we plot $C N^{-\alpha/\nu}$  vs $t-t_c$ for sum rule with $m=5$ shows 
collapse of all the peaks with $\alpha/\nu=0.5119(2)$. We obtain the same results for SR.
 (e) We plot $C N^{-\alpha/\nu}$ vs $(t-t_c)N^{1/\nu}$ using $1/\nu=0.519577$
we find that all the distinct  plots in (a) and (b) collapse into their own scaling curve.
 In (f) we plot $A_1C N^{-\alpha/\nu}$ vs $B_1(t-t_c)N^{1/\nu}$ and find that the scaling curves
 for different $m$ and rules collapse into a master curve for suitable metric factors $A_1$ and $B_1$.
 } 

\label{fig:4abcdef}
\end{figure}

Once we know the entropy, then 
finding the specific heat is just a routine matter since we know its general definition
\begin{equation}
C=({\rm control \ parameter}) \times {{d({\rm Entropy})}\over{d ({\rm control \ parameter})}}.
\end{equation}
In the case of thermal phase transition, we use temperature $T$ 
in the place of control parameter and Gibb's or Boltzmann entropy in the place of entropy. In the case of 
percolation, we use $q=(1-t)$ as the equivalent counterpart of temperature since both entropy and order
parameter behave with $q$ in the similar fashion as their corresponding counterpart in the
thermal phase transition. On the other hand, we use Shannon entropy as the equivalent 
counterpart of Gibb's or Boltzmann entropy. The definition of specific heat for 
percolation therefore is
\begin{equation}
\label{eq:specific_heat}
C(t)=q(t){{dH}\over{dq(t)}}.
\end{equation}
Thus, differentiating $H$ from first principles and multiplying the resulting value by the corresponding value of $q=(1-t)$,
we can immediately obtain $C(t)$. We then plot $C(t)$ in Figs. (\ref{fig:4a}) and (\ref{fig:4b}) as a function
of $t$ for PR and SR respectively and immediately see the sign of divergence at the critical point. Of course, the true divergence can only be
seen in the case  of the system size $N\rightarrow \infty$ which is an unsurmountable limitation of the numerical simulation.
One way of overcoming this is by using the finite-size scaling (FSS) theory. It helps us to extrapolate
the values of the critical exponents for infinite system from a set of data obtained from finite 
size system. We observe from Figs. (\ref{fig:4a}) and (\ref{fig:4b}) that
a given $C$ value occurs at an increasingly higher $t$ value as we increase the network size. It 
shows a sign of divergence at the critical point in the limit $N\rightarrow \infty$. To prove
this we can use the following finite-size scaling (FSS) hypothesis
\begin{equation}
\label{eq:fss_specific}
 C(t,N)\sim N^{\alpha/\nu}\phi_{C}((t-t_c)N^{1/\nu}),
\end{equation}
where $\phi_{C}(z)$ is the universal scaling function for specific heat. One way of proving this relation
is by data-collapse of distinct $C(t,N)$ versus $t$ for different $N$ into one universal curve $\phi_C(z)$ 
which is our next task.

According to Eq. (\ref{eq:fss_specific})  the peak $C_{\rm h}$ of the $C(t,N)$ vs $t$ curves
at $t=t_c$ increases following a power-law
$C_{{\rm h}}\sim N^{\alpha/\nu}$. Plotting  $\log(C_{{\rm h}})$ versus $\log(N)$ in Fig. (\ref{fig:4c}) we find a 
 set of parallel lines for different $m>1$ and for different rules whose slope 
gives an estimate of $\alpha/\nu$. The same value of $\alpha/\nu$ for all $m$ and product-sum rules
suggests that they all share the same critical exponents. 
Plotting $C(t,N)N^{-\alpha/\nu}$ versus $t$, see Fig. (\ref{fig:4d}), and finding that all the peaks collapse 
to one point provides a clear testament of how good is this value. We checked 
it for $m=3,5,10$ and their product and sum rules. We find that all the peaks collapse at the same point 
with the same value of $\alpha/\nu=0.5119$.
We then draw a horizontal line in the plot of $C(t,N)N^{-\alpha/\nu}$ versus $t$ and measure
the distance of the intercepts of each curve from the critical point to obtain a data 
of $t-t_c$ versus $N$ \cite{ref.hassan_didar, ref.hassan_sabbir}. Plotting this data in the $\log-\log$ 
scale gives a straight line whose
slope gives a good estimate for  $1/\nu$ value. Finally, we plot $CN^{-\alpha/\nu}$ 
vs $(t-t_c)N^{1/\nu}$  and after fine tuning the $1/\nu$ we obtain 
a perfect data-collapse with $\alpha=0.98739$ and $1/\nu=0.519577$  for both the rules as shown in Figs. (\ref{fig:4e}). Furthermore we plot  $A_1CN^{-\alpha/\nu}$ 
vs $B_1(t-t_c)N^{1/\nu}$ in Fig (\ref{fig:4f}) with
$m=3,5,10$  for both the rules and find that all the scaling curves 
collapse into one master curve once we use appropriate metric factors $A_1$ and $B_1$  \cite{ref.hsu}.  
We then use the relation  $N\sim (t-t_c)^{-\nu}$ in $C(t)\sim N^{\alpha/\nu}$ and 
find that indeed the specific heat diverges like
\begin{equation}
C(t)\sim (t-t_c)^{-\alpha}.
\end{equation} 
The quality of data-collapse is a clear testament of  the accuracy of our $\alpha=0.98739$ value. Note
that the $\alpha$ value of EP in BA network is slightly lower than its value in the ER network
where $\alpha=1$.

\begin{figure}

\centering
\subfloat[]
{
\includegraphics[height=2.4 cm, width=4.00 cm, clip=true]
{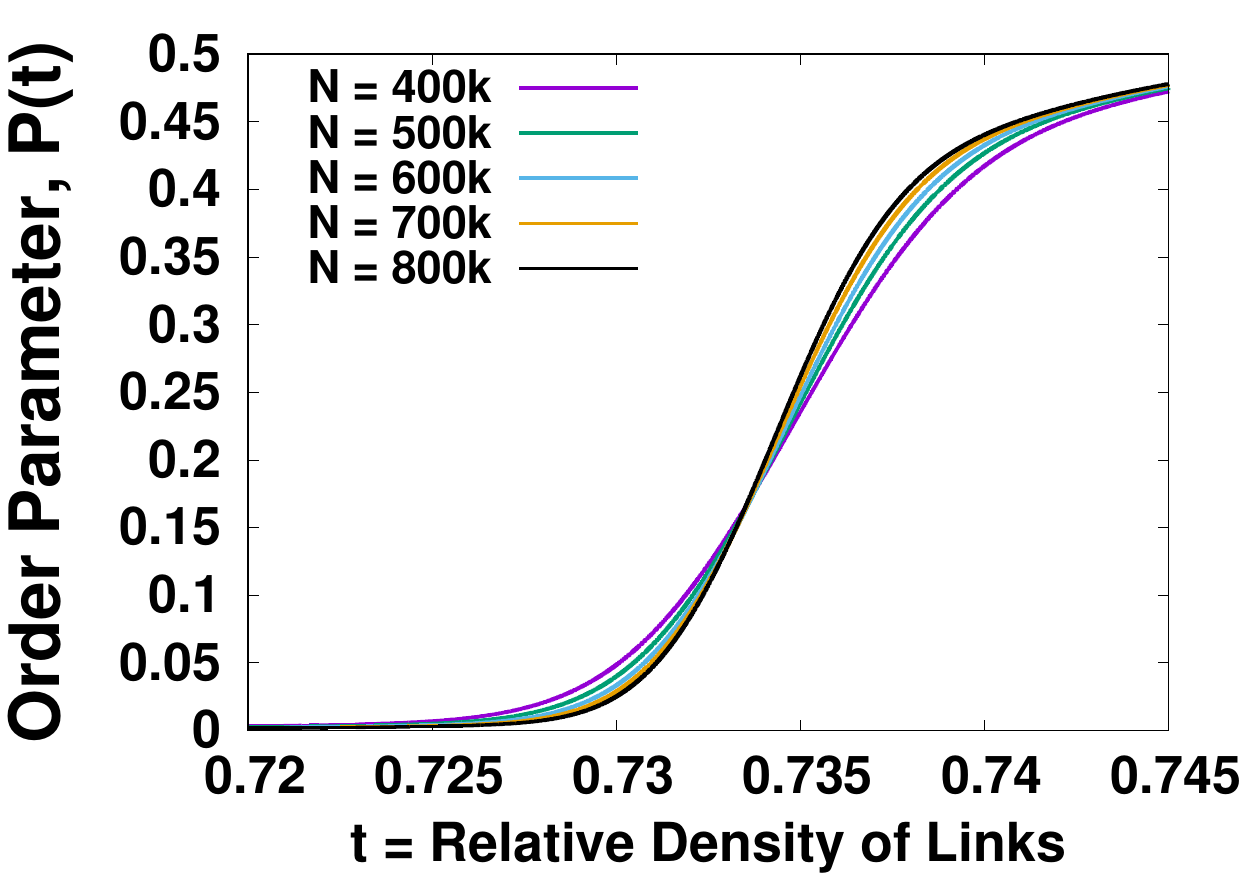}
\label{fig:5a}
}
\subfloat[]
{
\includegraphics[height=2.4 cm, width=4.00 cm, clip=true]
{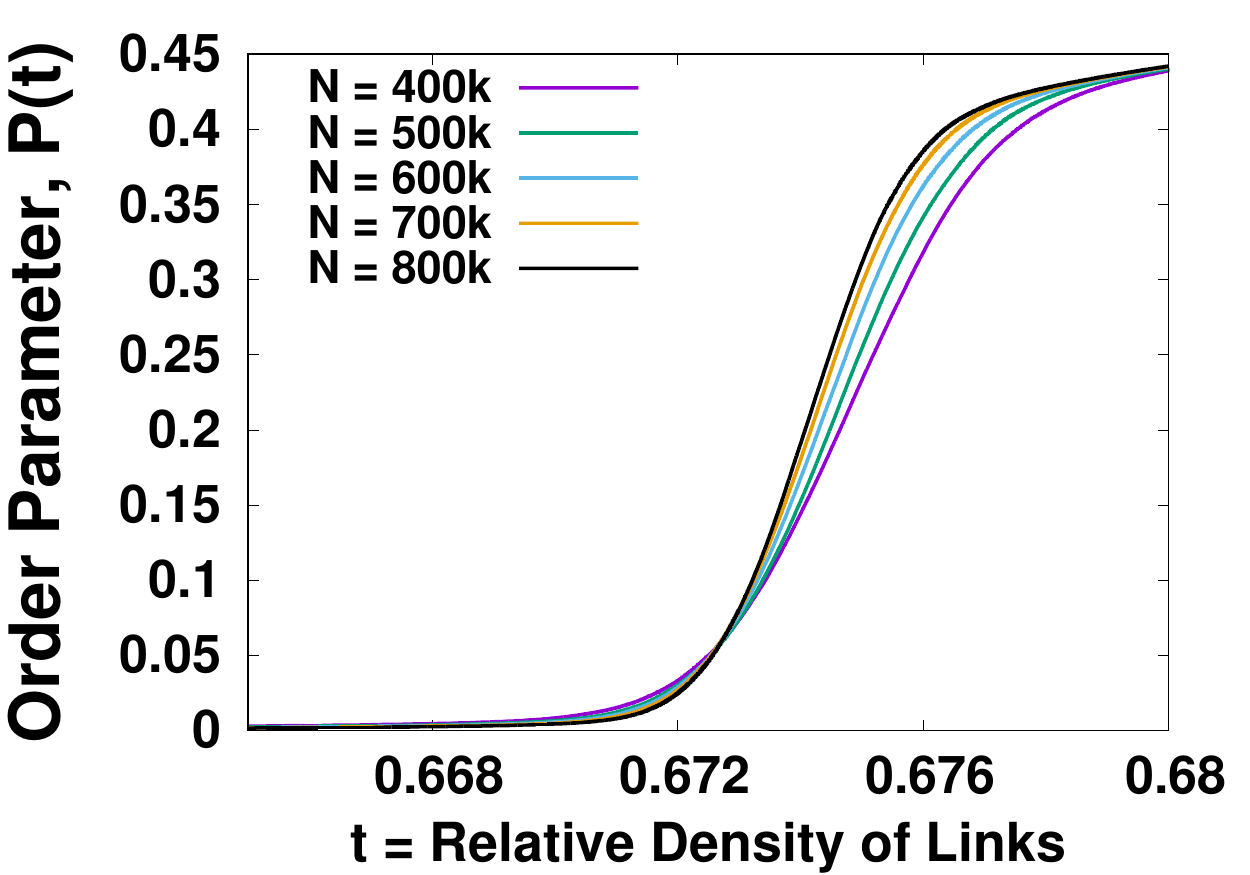}
\label{fig:5b}
}


\subfloat[]
{
\includegraphics[height=2.4 cm, width=4.00 cm, clip=true]
{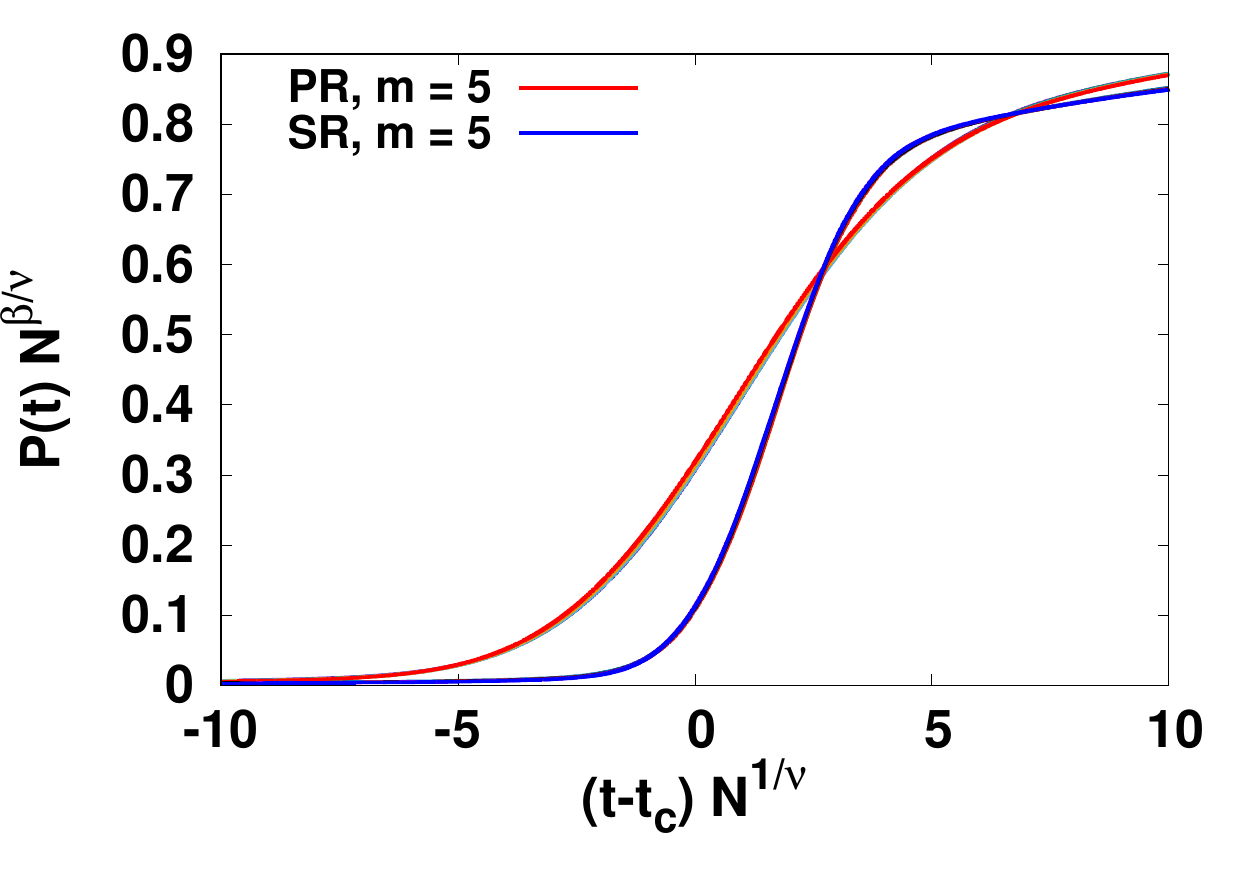}
\label{fig:5c}
}
\subfloat[]
{
\includegraphics[height=2.4 cm, width=4.00 cm, clip=true]
{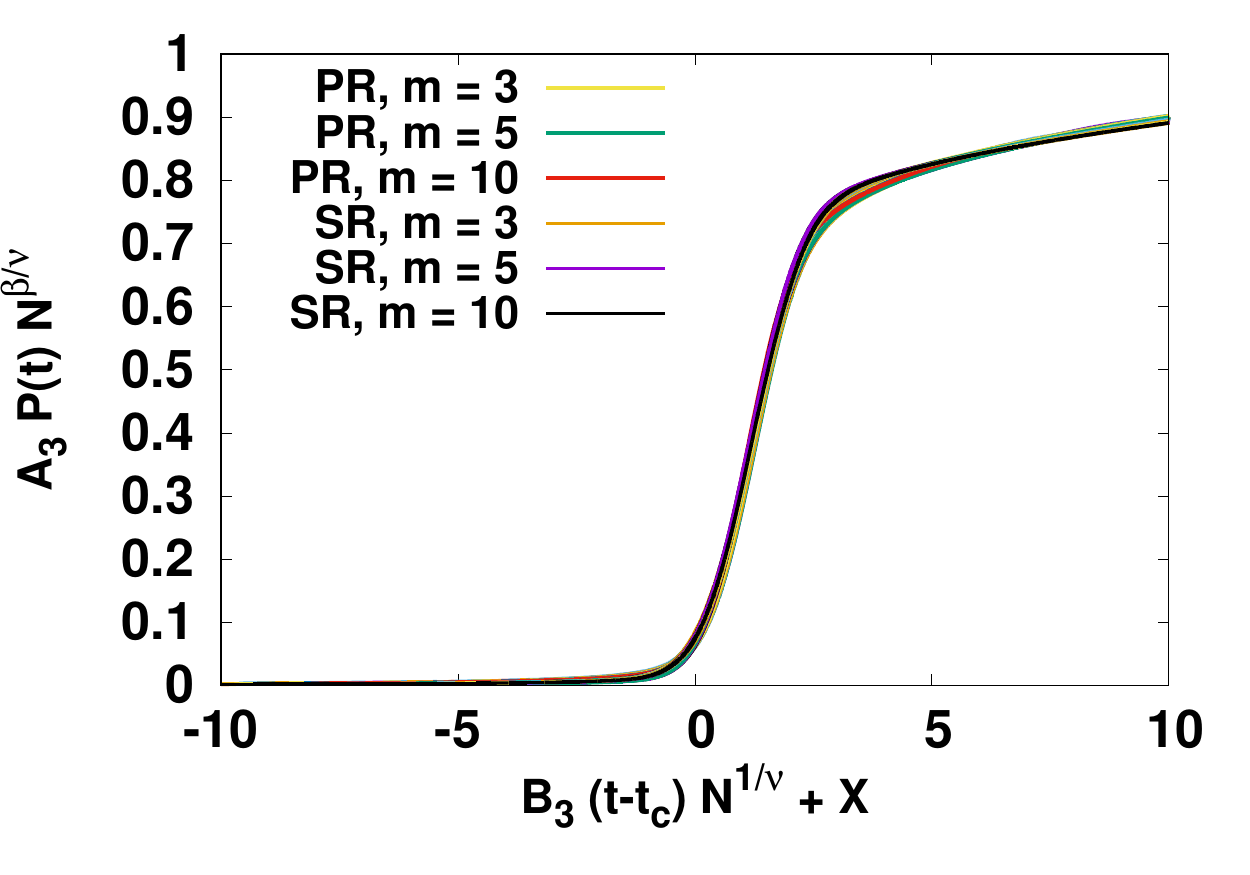}
\label{fig:5d}
}
\caption{
We show plots of $P$ versus $t$ (a) for both PR and (b) for SR.  In (c) 
we plot $PN^{\beta/\nu}$ versus $(t-t_c)L^{1/\nu}$ for both PR and SR. We
find that all the distinct plots in (a) and (b) collapse into their own scaling curves sharing the same critical exponents. (d) Plots of $A_2PN^{\beta/\nu}$ versus $B_2(t-t_c)L^{1/\nu}$
for PR-SR collapse into a master curve if we can choose
suitable values for metric factors $A_2$ and $B_2$ independent of $m>1$. 
} 

\label{fig:5abcd}
\end{figure}

Our next goal is to find the critical exponent $\beta$ of the order parameter. 
Essentially, we should find the $\beta$ value also for infinite system for which $P$ should
be equal to zero over the entire regime of $t<t_c$. This is, however, not the case 
as long as we work with finite size system. Plotting $P(t)$ as a function of $t$ for different system
size clearly shows a sign that as $N$ increases we find $P=0$ up to an increasingly higher 
value of $t$ as shown in Figs. (\ref{fig:5a}) and (\ref{fig:5b}).  The order parameter is said to
exhibit finite-size scaling if it can be expressed as
\begin{equation}
P(t,N)\sim N^{-\beta/\nu}\phi_\beta((t-t_c)N^{1/\nu}),
\end{equation}
where $\phi_\beta(x)$ is the universal scaling function of $P$. Note that unlike the finite size
scaling ansatz for $C(t, N)$, we have a negative sign in the exponent for
order parameter. It implies that if we plot $P(t,N)$ vs $z=(t-t_c,N)N^{1/\nu}$ then the value
of $P$  decreases with network size. By measuring the extent of decrease at a fixed value of $z$ 
somewhere at $z>0$, we collect its data as a function of $N$. Plotting the resulting data in the log-log scale 
gives us a straight line whose slope is just a rough estimate of the
value $\beta/\nu$. All the  distinct curves of Figs. (\ref{fig:5a}) and (\ref{fig:5b})
collapse into a universal curve as shown in Fig. (\ref{fig:5c}) 
if we plot  $PN^{\beta/\nu}$ vs $(t-t_c)N^{1/\nu}$ with
$\beta/\nu=0.0468$ and $1/\nu=0.5195$ regardless of whether it is PR or SR or of the $m$ value. 
In Fig. (\ref{fig:5d}) we plot $A_2PN^{\beta/\nu}$ vs $B_2(t-t_c)N^{1/\nu}-X$
for PR-SR with different $m$ and find that their respective
scaling curve collapse into a master curve sharing the same critical exponents. It implies that their
scaling functions are the same except for a multiplying metric factors $A_2$, $B_2$ and trivial
shifting factor $X$.
Substituting the relation $N\sim (t-t_c)^{-\nu}$ in $P\sim N^{-\beta/\nu}$ we get
\begin{equation}
P(t)\sim (t-t_c)^\beta.
\end{equation}
This is exactly how the order parameter behaves near critical point in the thermal CPT as well. 
We once again find that both PR and SR rules for all $m>1$ share the same
exponent $\beta =0.09007$ which is slightly higher than that in the 
ER network.

\begin{figure}

\centering

\subfloat[]
{
\includegraphics[height=2.4 cm, width=4.00 cm, clip=true]
{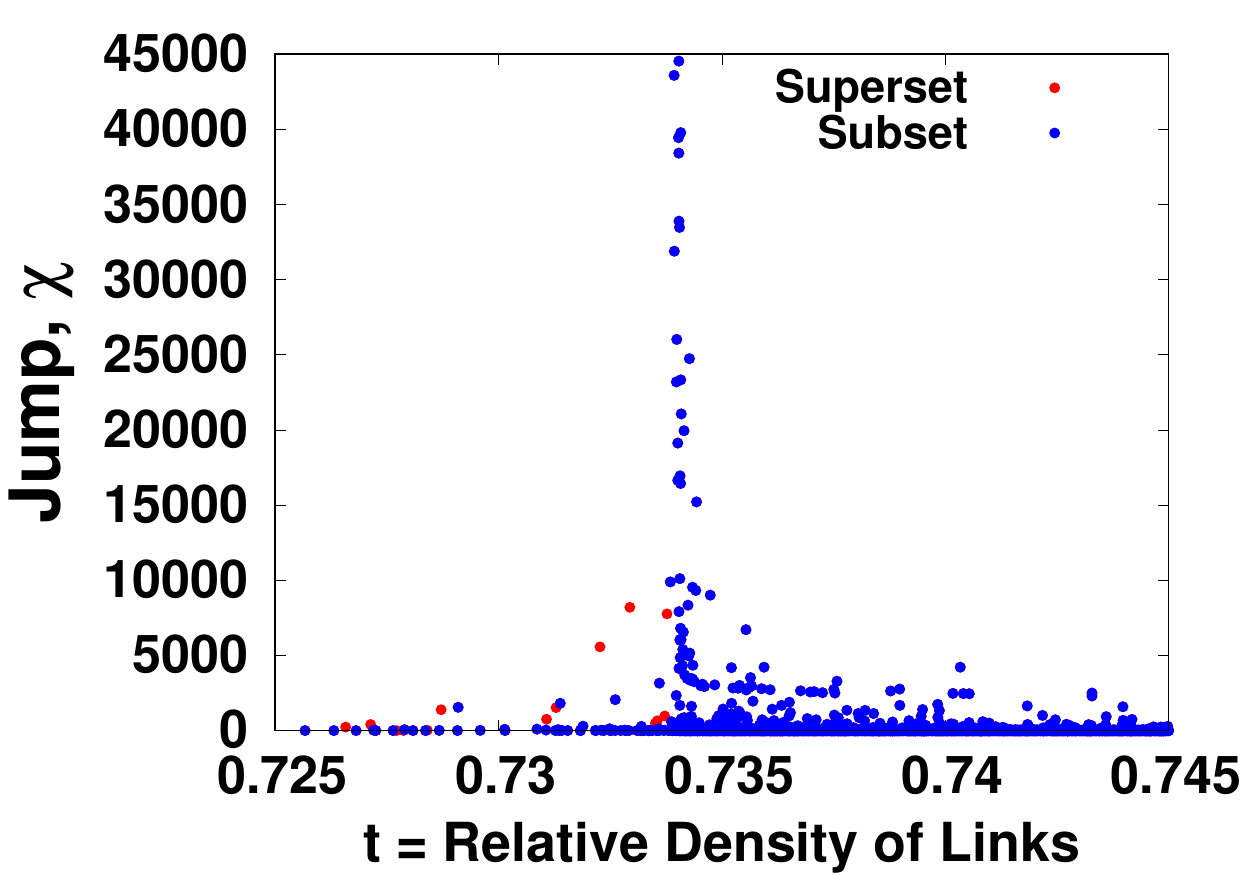}
\label{fig:6a}
}
\subfloat[]
{
\includegraphics[height=2.4 cm, width=4.00 cm, clip=true]
{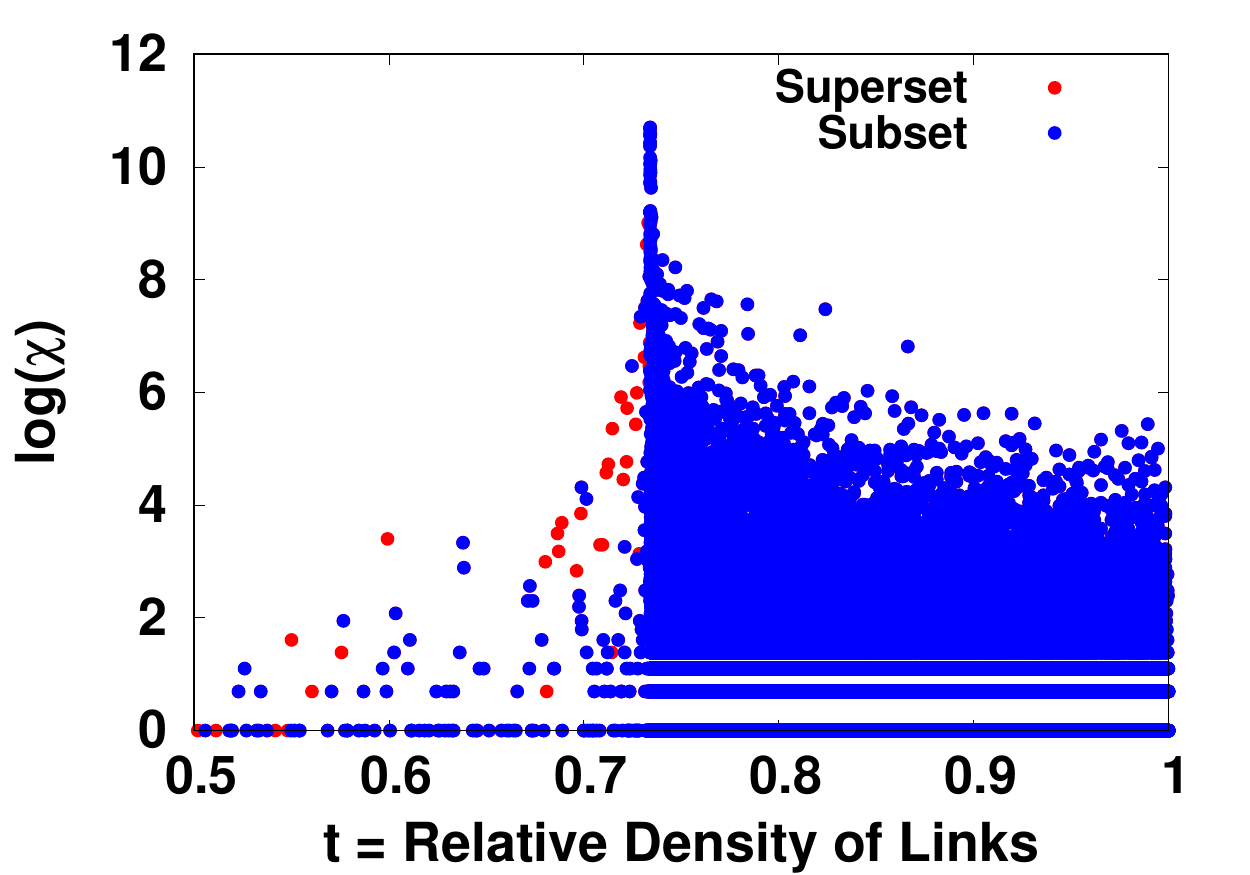}
\label{fig:6b}
}

\caption{
(a) The jump size in the largest cluster is plotted against $t$ 
using data for single realization. A subset of the same data point (blue dots)
which corresponds to those jumps which occur only to the existing largest clusters
is plotted again in the same plot.  (b) The same data is shown in the $\log$-linear scale. 
} 
\label{fig:6ab}
\end{figure}

Recently, we defined the susceptibility $\chi(t,N)$ for percolation as the ratio of the change
in the order parameter $\Delta P$  and the magnitude of the time interval $\Delta t$ during which 
the change $\Delta P$  occurs. Essentially it becomes the derivative of the order parameter $P$ since
$\Delta t\rightarrow 0$ in the limit  $N\rightarrow \infty$ as $\Delta t=1/N$. 
Recall that susceptibility in the paramagnetic 
to ferromagnetic transition too is the derivative of the order parameter.
Using $\Delta P=\Delta s_{{\rm max}}/N$ in the definition we find 
\begin{equation}
\chi(t,N)=\Delta s_{{\rm max}},
\end{equation}
which is essentially the jump in the largest cluster as we add link one by one. 
Note that it is not always guaranteed that each time we add a link 
a jump in the largest cluster $s_{{\rm max}}$ will occur. There are
two distinct ways in which jump can takes place. It may happen that an already existing largest
cluster $s_{{\rm max}}^i$
may merge with a smaller cluster whose combined size $s_{{\rm max}}^f$ is obviously 
larger than $s_{{\rm max}}^i$. Besides, it
may also happen that two smaller clusters may merge such that their
 combined size $s_{{\rm max}}^f$ may be larger than $s_{{\rm max}}^i$.
In either case the jump in the largest is the difference between  $s_{{\rm max}}^f$ and 
$s_{{\rm max}}^i$. In order to understand the fine details of what happens as we keep
adding links let us see which happens more or less frequently compared to the other 
graph with red dots. To appreciate that
we plot the susceptibility  $\chi(t,N)=\Delta s_{{\rm max}}$
as a function of $t=n/N$ using data from a 
single realization which is shown in Fig. (\ref{fig:6a}) represented by the  red dots. We re-plot 
a subset of the above data for whom the jumps 
in $s_{{\rm max}}$ occurred only to the existing largest cluster on the same graph with red dots.
We see there is no red dots at above $t_c$.
 To appreciate it even better we plot
$\log (\Delta s_{{\rm max}})$ versus $t$ in Fig. (\ref{fig:6b}) \cite{ref.manna}. It clearly shows that there is 
a sharp demarcation line at the critical point that distinguishes two regimes. In the regime $t<t_c$,  
there are more red dots than blue> It implies that in this regime, the smaller clusters join more frequently to
undergoes a jump over the existing largest cluster than jump due to growth of an
already existing largest cluster by joining with a smaller cluster.  On the other hand,   
in the regime $t>t_c$, there are only red dots signifying that 
there exists a unique largest cluster as jump always occurs to an already existing one.

\begin{figure}

\centering

\subfloat[]
{
\includegraphics[height=2.4 cm, width=4.00 cm, clip=true]
{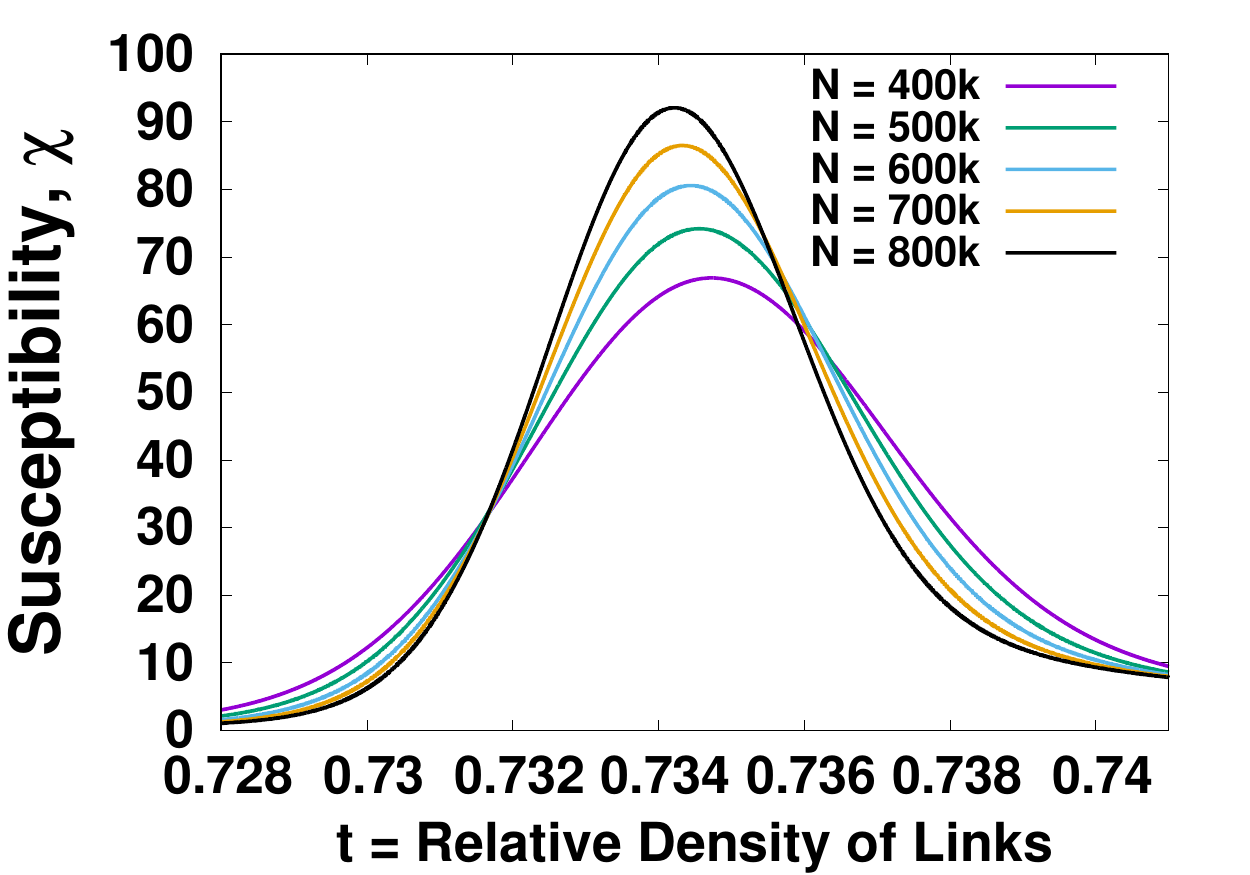}
\label{fig:7a}
}
\subfloat[]
{
\includegraphics[height=2.4 cm, width=4.00 cm, clip=true]
{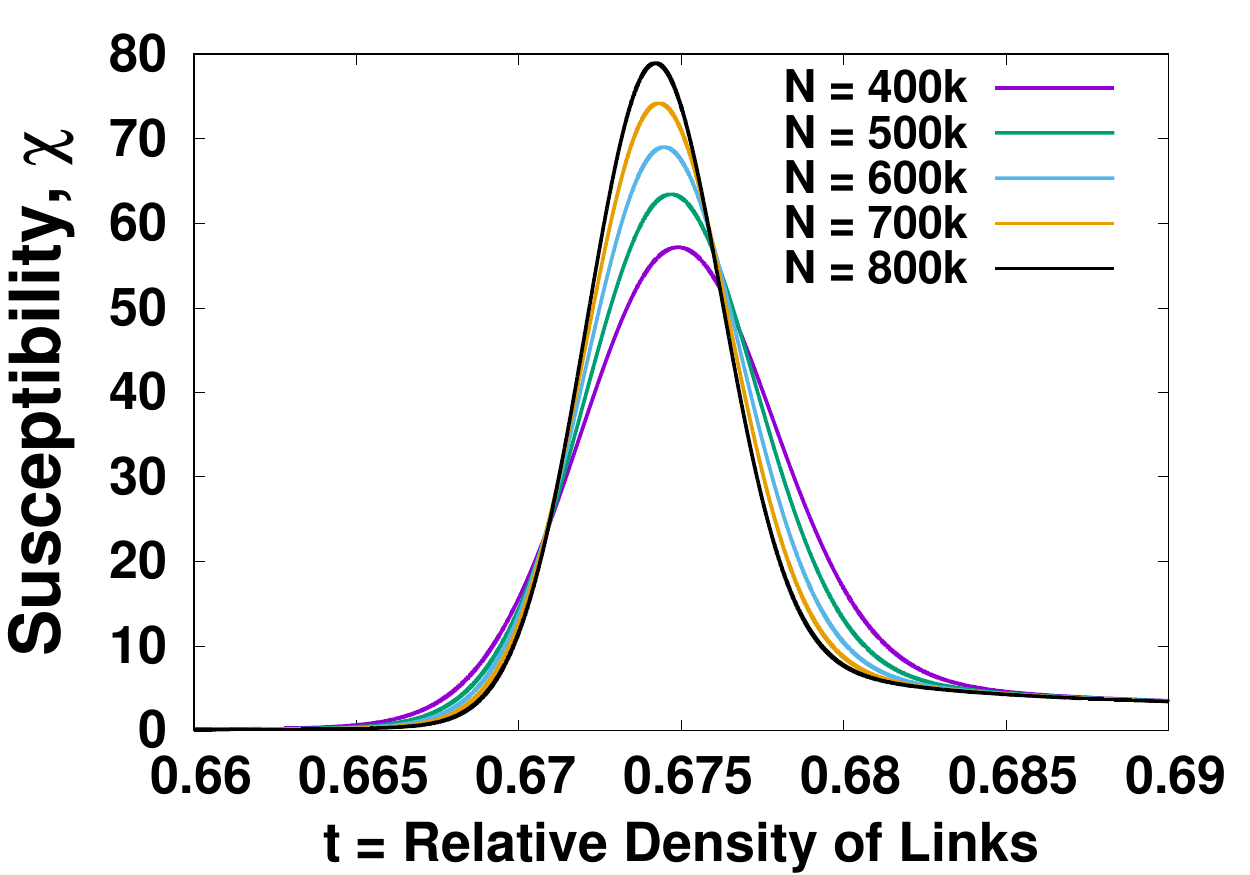}
\label{fig:7b}
}

\subfloat[]
{
\includegraphics[height=2.4 cm, width=4.00 cm, clip=true]
{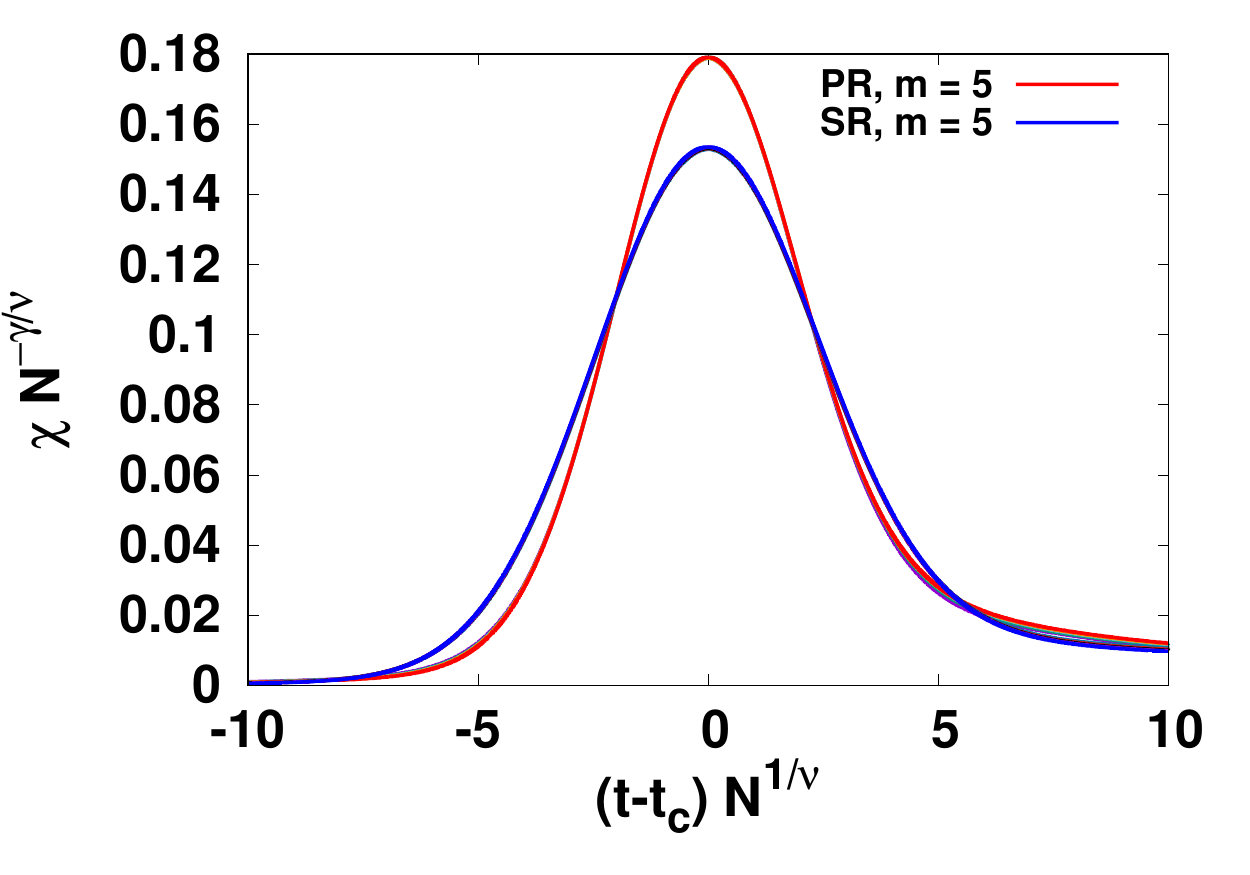}
\label{fig:7c}
}
\subfloat[]
{
\includegraphics[height=2.4 cm, width=4.00 cm, clip=true]
{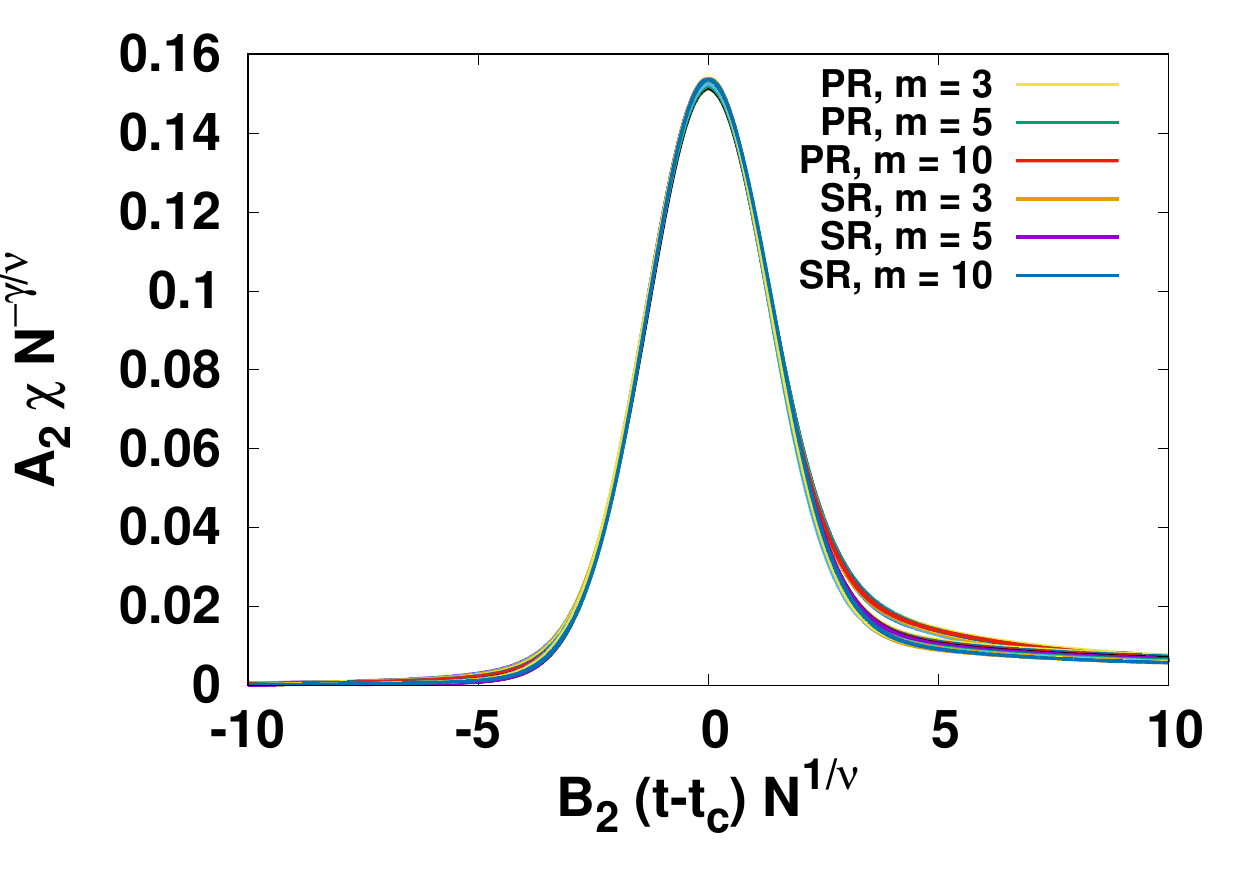}
\label{fig:7d}
}

\caption{
Plots of susceptibility $\chi$ versus $t$ with $m=5$ (a) for PR and (b) for SR. 
(c)  We plot $\chi N^{-\gamma/\nu}$ vs $(t-t_c)N^{1/\nu}$ and find distinct curves in (a) and (b)
collapse into their own scaling curves. (d) We plot the same for PR-SR with different $m$ and that
distinct scaling curves collapse into a master curve if we can choose suitable metric factors $A_3$ and $B_3$.
} 

\label{fig:7abcd}
\end{figure}

Percolation theory is a statistical model and hence the numerical values of an observable 
from a single realization are not reproducible. We thus take the average of $\chi$
value over many independent realizations $M$ under the same condition and find that the
resulting plot  looks apparently different from the ones for single realization albeit the key 
features are the same. Besides, if we 
take another set of ensemble average over the same number of independent realizations under
the same condition then the numerical value will be identical at least in the limit $M\rightarrow \infty$ and it will be increasingly
more so as $N\rightarrow \infty$. In Figs. (\ref{fig:7a}) and (\ref{fig:7b}) 
we plot ensemble average of $\chi$ as a function of $t$. This leads to assume that it too
should obey the finite-size scaling form
\begin{equation}
\label{eq:4}
 \chi(t-t_c,N)\sim N^{\gamma/\nu}\phi_\chi((t-t_c)N^{1/\nu}).
\end{equation}
where $\phi_\chi(\xi)$ is known as the universal scaling function. 
In Figs. (\ref{fig:7a}) and (\ref{fig:7b}) we show plots 
of $\chi$ versus $t$ for both 
PR and SR model. According to Eq. (\ref{eq:4}), the susceptibility  $\chi_{{\rm max}}$ 
at $t=t_c$ increases obeying a power-law $\chi_h \sim N^{\gamma/\nu}$. Following the same procedure as
we have done for specific heat we find $\gamma/\nu=0.459(2)$ for PR and $\gamma/\nu=0.458(4)$ 
for SR. 
We now plot  $\chi N^{\gamma/\nu}$ vs $(t-t_c)N^{1/\nu}$ in Fig. (\ref{fig:7c}) and 
find that all the distinct curves in Figs.
(\ref{fig:7a}) and (\ref{fig:7b}) collapse superbly into their respective scaling curves
if we use $\gamma/\nu=0.4594$ and $1/\nu=0.5195$ for both PR
and SR. Note that $t_c$ value also affects the data collapse and hence, tuning the initial 
estimates for $t_c$, we get the best data-collapse if we use 
$t_c=0.73433$ for PR and $t_c= 0.67420$ for SR. In Fig. (\ref{fig:7c})
we plot $A_3\chi N^{\gamma/\nu}$ vs $B_3(t-t_c)N^{1/\nu}$ and find that PR-SR belong to the
same universality class regardless of the value of $m>1$.
 Using now the relation $N\sim (t-t_c)^{-\nu}$ in $\chi\sim N^{\gamma/\nu}$ 
we find that 
\begin{equation}
\chi\sim (t-t_c)^{-\gamma},
\end{equation}
where $\gamma=0.91796$ for both PR and SR within the acceptable limit of error. 
It clearly shows that the susceptibility now diverges even without the exclusion of the largest cluster and that too
with the same $\gamma$ value for both the rules.



Scaling theory predicts that the various critical exponents cannot just assume values independently 
rather they are bound by some scaling and hyperscaling relations.  It has been known that the exponents $\alpha$, $2\beta$ and $\gamma$ 
describing specific heat, square of order parameter and susceptibility are related by 
Rushbrooke inequality $\alpha+2\beta+\gamma=2+\epsilon$ where $\epsilon>0$ but negligibly small.
Remarkably, many experiments and exactly solved models of thermal CPT suggest that 
the Rushbrooke equality rather holds more as equality than as inequality. The static Widom scaling (SWS) 
also suggests such equality. The basic assumption of the SWS is that the Gibb's free energy 
for magnetic system is a generalized homogeneous function i.e., 
\begin{equation}
G(\lambda^{a}\epsilon,\lambda^{b} h)=\lambda G(\epsilon,h), \hspace{0.25cm} \forall \ \lambda,
\end{equation}
 where $h$ is the external magnetic field  and $a$ and $b$ are two scaling parameters \cite{ref.Stanley}. 
If we now use the definitions of specific heat, square of magnetization and susceptibility 
then we find that near the critical point they exhibit power-law and the corresponding critical exponents are
\begin{equation}
2\beta=2{{1-b}\over{a}}, \hspace{0.25cm} \alpha=2-{{1}\over{a}} \hspace{0.25cm}
{\rm and} \hspace{0.25cm} \gamma={{2b-1}\over{a}}.
\end{equation}
Summing these relations we obtain the Rushbrooke inequality 
in the form of equality. Substituting our values of $\alpha=0.98739$,  $\beta =0.09007$ and $\gamma=0.91796$ in
the Rushbrooke relation we find 
\begin{equation}
\alpha+2\beta+\gamma=2.085. 
\end{equation}
It clearly suggests that, within the limits of error, Rushbrooke equality holds in explosive percolation.
Earlier we have found the same result for EP in ER network, RP in square and weighted planar stochastic lattice \cite{ref.hassan_sabbir, ref.Hassan_Rahman_1}.

The classifications of second order thermal phase transition under different physical conditions into universality classes
is yet another interesting proposition.  It has been put forth by Kadanoff in 1970 who suggested that two systems with the same values of 
critical-point exponents and scaling functions are said to belong to the same universality class. It has been found
that many different materials near their own critical points scales sharing the same critical exponents provided they have the
same spatial and spin dimension, the same range of interaction but strength of interactions mat vary. The similar results are also found
in random percolation.  It is well-known that RP in all lattices
having the same spatial dimensions belong to the same universality class regardless of whether it is bond or site type except one exception. Recently
we have shown that RP in a scale-free multifractal lattice does not belong to the universality class of other two dimensional lattices albeit its spatial 
dimension is the same as square lattice \cite{ref.hassan_didar, ref.Hassan_Rahman_1}. Explosive percolation has been introduced only nine years ago while random percolation
was introduced more than sixty years ago. Since then much of the most active time of those nine years have been spent in 
resolving the issue of whether EP describes continuous  or discontinuous phase transition. There
has hardly been any attempt to find the critical exponents for EP under different conditions \cite{ref.Hassan_Rahman_explosive}. 
Besides, scope for checking Rusbrooke inequality has only been opened up after our definitions for entropy, specific heat and susceptibility
 \cite{ref.hassan_didar}.   Earlier Bastas {\it et al.} reported
that site and bond type explosive percolation on the same lattice do not belong to the same universality 
class \cite{ref.Bastas}.
 Radicchi and Castellano on the other hand reported that 
the site-bond universality is violated even in the case of random percolation if the substrate is a network with null percolation thresholds \cite{ref.Radicchi_Castellano}. 
However, there do not exist much works in network to see whether 
whether random or explosive percolation in network too can be classified into universality classes. 
The classifications of explosive percolation into 
universality classes is far from complete. Only recently, we have shown that explosive percolation of PR and SR type in the ER network belong to the
same universality class.However, we have recently shown that product and sum rules of the explosive percolation
in ER network belong to the same universality class \cite{ref.hassan_sabbir}. The present work can be considered as yet another
development towards classifying EP into universality classes. 
Finding that 
PR and SR of explosive percolation in the scale-free BA network belong to the same universality class 
regardless of the value of $m$ except $m=1$ is indeed a significant development.

To summarize, we have used a class of BA networks, depending on the value of $m$, as a skeleton
to study explosive percolation for both product and sum rule of the Achlioptas process. 
Our goal was to investigate the role of $m$ as well as of the product and sum rule in fixing 
the critical points and the critical exponents. Our second goal was to check if the entropy $H$, that measures the degree of 
disorder, is consistent with the behaviour of the order parameter $P$, that measures the extent of order. To this end, we find that 
behaviour is reminiscent of what we see in its thermal counterpart such as in the paramagnetic
to ferromagnetic transition. It suggests that explosive percolation is an order-disorder transition except for $m=1$
for which we do not have access to the phase $t>t_c$ like one dimensional percolation.
However, for $m>1$ we have found non-trivial $t_c$ value  which 
decreases systematically with increasing $m$. We have also found the
critical exponents $\alpha, \beta, \gamma$ and $\nu$ of the equivalent counterpart of 
specific heat, order parameter, susceptibility and correlation length respectively. Interestingly, we found that
these values are independent of  $m$ and of the rules albeit $t_c$ values are different. It implies that
there is one unique  universality class for EP in the whole class of BA networks provided $m>1$ .  The fact that the critical exponents and the scaling
functions are the same apart from scale factors for all $m>1$ regardless of rules, namely product and sum rules, is truly remarkable.
 It is well-known in thermal continuous phase transition that the exponents $\alpha$, $2\beta$ and $\gamma$ 
describing specific heat, square of order parameter and susceptibility are related by the simple scaling law
$\alpha+2\beta+\gamma=2+\epsilon$ where $\epsilon>0$ but negligibly small. This has been found true in thermal continuous phase transition as well as
the RP.
Indeed, we have shown that this is also true for explosive percolation where the value of the critical exponents $\alpha$, $\beta$ and $\gamma$ 
obeys the Rushbrooke inequality. We believe that the present work
will be of great interest to the community.

\end{document}